\documentclass{frontiersSCNS} 
\usepackage{url,hyperref,lineno,microtype}
\usepackage[onehalfspacing]{setspace}
\linenumbers
\pdfoutput=1

\def\keyFont{\fontsize{8}{11}\helveticabold }
\def\firstAuthorLast{James Kozloski} 
\def\Authors{James Kozloski\,$^{1,*}$}
\def\Address{$^{1}$Computational Biology Center, IBM Research Division, IBM T.J. Watson Research Center, Yorktown Heights, New York, U.S.A.}
\def\corrAuthor{James Kozloski}
\def\corrAddress{Computational Biology Center, IBM Research Division, IBM T.J. Watson Research Center, Yorktown Heights, New York, 10598, U.S.A.}
\def\corrEmail{kozloski@us.ibm.com}

\begin{document}
\begin{nolinenumbers}
\onecolumn
\firstpage{1}

\title[Brain Model of Information Based Exchange]{Brain Model of Information Based Exchange}

\author[\firstAuthorLast ]{\Authors} 
\address{} 
\correspondance{} 

\extraAuth{}

\maketitle

\begin{abstract}
\section{}
Here we describe an ``information based exchange'' model of brain function that ascribes to neocortex, basal ganglia, and thalamus distinct network functions. The model allows us to analyze whole brain system set point measures, such as the rate and heterogeneity of transitions in striatum and neocortex, in the context of disease perturbations. Our closed-loop model invokes different forms of plasticity at specific tissue interfaces and their principle cell synapses to achieve these transitions. By modulating information based exchange of action potentials between modeled neocortical areas, we observe changes to these measures in simulation. We hypothesize that similar dynamic set points and modulations exist in the brain's resting state activity, and that germ line modifications of information based exchange may increase the risk of diseases such as Huntington's, Parkinson's, and Alzheimer's. Disturbances in synaptic plasticity at distinct tissue interfaces in the model may be used to estimate risks of system dysfunction and neuronal cell death from quantitative analyses of the global dynamics that maintain system set points. The model is targeted for further development using IBM's Neural Tissue Simulator, which allows scalable elaboration of networks, tissues, and their neural and synaptic components towards ever greater complexity and biological realism. Elaboration of these simulations within each modeled neural tissue allows \emph{in silico} study of therapeutic interventions in living brain tissue.

\tiny
 \keyFont{ \section{Keywords:} neocortex thalamus basal ganglia information based exchange brain model} 
\end{abstract}

\section{Introduction}

Synaptic plasticity regulates neuronal responses to patterns of inputs impinging on dendritic arbors from multiple presynaptic sources. Resulting input selectivity is often associated with learning and memory: cognitive functions often reduced at the single neuron level to simple input-output categorizations. At the circuit level, synaptic plasticity can serve more complex functions over arbitrary inputs, from selecting fixed points in recurrent networks \citep{Hopfield1982}, to implementing optimizations such as information maximization in artificial neural networks \citep{Linsker1997}, to dynamical encoding by winnerless networks \citep{Rabinovich2001}. A challenge to analyzing the role of any neuron or circuit that implements these functions for cognition is that of modeling appropriate, naturalistic neuronal and circuit inputs, which in real brains derive from tens of thousands to millions of other neurons.

Here we present a closed-loop brain model, including component models of several neural tissues that we hypothesize implement some of these functions. Synapses and plasticity connecting components at principle cell interfaces together create a set of closed neuroanatomical loops. Without extrinsic inputs or stochastic intrinsic drivers, our model avoids the challenges and assumptions of modeling naturalistic inputs separately, and instead derives them exclusively from the dynamics of upstream neurons and tissues within the model's closed loops. The challenge then is model validation, which we won't address in this report. Instead the aim here is to delineate hypotheses and a theory of brain resting state function using the model and its simulations. We propose that models implemented similarly constitute a class of ``brain models,'' and are distinct from component ``neural tissue models,'' which instead must assume a set of inputs or stochastic processes to drive intrinsic tissue dynamics. In this way, a coarse but consistent model of global brain function may be useful for constraining the most detailed neural tissue simulations.

We introduce the term ``traversal'' to refer to a ``synfire chain'' as defined by Abeles \citep{Abeles}, but with additional neuroanatomical constraints defining a minimum set of neocortical regions traversed by the event. The cortico-cortical feedback loop in our model acts as a substrate for combined traversals of sensory, limbic, and motor areas, which we propose together drive behavior in the organism. The cortico-thalamo-cortical feed forward loop acts to maximize the entropy of these global traversals and to maximize information about the environment relayed as inputs to the loop. Lastly, the striato-nigro-striatal loop provides a means to select subsequent configurations by monitoring changes in ongoing traversals and signaling them with dopamine to alter routing within the feed forward entropy maximizing network. We propose this function as the substrate for reward learning in the organism. Each of these loops therefore has both a closed-loop function (global traversal, traversal entropy maximization, and traversal change monitoring and rerouting) and an organismal input-output function (behavior generation, sensory processing, and behavior selection based on reward learning).

The objective of this report is to describe the closed-loop model and simulations of it, and how we hypothesize about dynamic disease mechanisms and progression based on them. Furthermore, methods for modeling treatments that alter brain disease risks using neural tissue simulation \citep{Kozloski2011} and perturbations to the closed loop that alter dynamic set points will also be described. To summarize our overall approach, the driving hypotheses for brain disorder and disease that we aim to delineate using the model are: $1.$ The primary disease and disorder risk is a disturbance in plasticity that critically maintains brain system dynamic set points; $2.$ Compensatory circuit dynamics achieves near-normal set points despite genetic or environmental perturbations, but with increased secondary risk of neuronal dysfunction, damage, or loss; $3.$ Secondary risk correlates with feed forward destruction or dysfunction of neural tissues because with each neuron function lost, maintaining system set points requires even greater secondary risks; and $4.$ Slowing progression may therefore lie in mitigating the primary risk's effect on system set points or in limiting secondary risks incurred by inherent compensatory dynamics.

\section{Information Based Exchange Brain Model}

\subsection{Cytoarchitectonics of Bidirectional Neocortical Projections: The ``Grand Loop''}

We propose a model that emphasizes a specific cortico-cortical connectivity across the major sensory, limbic, and motor categories of Brodmann areas. This emphasis derives from several observations. First, we note the importance of signals traversing all three categories of cortical representations in order to produce a stable basis for perception and behavior by integrating information about the environment, internal needs, and behavioral opportunities of the organism. While many loops have been discovered in studies of the neocortical connectome, none provide the directed graph (feed forward vs. feedback) needed to identify a system to support such traversals. Instead, we note that the cytoarchitectonic granularity of neocortical areas provides one means to interpret feed forward (i.e., more granular to less granular) and feedback (i.e., less granular to more granular) connections between cortical areas \citep{Rempel-Clower2000} and therefore a means to identify a backbone for global brain traversals (Fig. \ref{Loop}A).

Granularity refers to the density of punctate Nissl bodies in stained layer $4$ of neocortex. The granularity across all of neocortex was studied and mapped extensively by von Economo \citep{vonEconomo}, and we reproduce his illustrations and some key findings in Figures \ref{Loop}A and \ref{Loop}B. Note that granular cortices typically have smaller diffuse Nissl bodies in layer $5$, and agranular cortices have very large diffuse layer $5$ Nissl bodies. Tiling in von Economo's map shows that regions of cortex with similar granularity are adjacent, with key exceptions at the boundaries between primary motor (M1) and primary somatosensory (S1), hippocampus (HC) and retrosplenial granular areas (RGA), and subgenual anterior cingulate (ACC) and prefrontal (PFC) cortices. Each of these three pairs of Brodmann areas are interconnected, and in our model represent key boundaries in the backbone for traversing the sensory-limbic (HC-RGA), limbic-motor (ACC-PFC), and motor-sensory (M1-S1) cortices (Fig. \ref{Loop}B, black arrows). To complete a ``Grand Loop'' backbone, we join each pair of areas by an area in their adjacent dysgranular neocortical regions: the secondary somatosensory (S2), posterior cingulate (PCC), and supplemental motor (SMA) areas (Fig. \ref{Loop}C). While others have noted that organizing principles for intrinsic microcircuits may be derived from combining von Economo's observations with those regarding granularity and the direction of cortico-cortical projections \citep{Beul2015}, none to our knowledge have proposed a Grand Loop that traverses all of neocortex according to these principles.

\subsection{Cortico-Cortical and Cortico-Thalamo-Cortical Functional Pathways}

Having defined the feed forward neocortical Grand Loop, we'll now embellish this structural model with additional components based on observations regarding feed forward projections and signaling between neocortical areas. Sherman and Guillery emphasized different roles for direct cortico-cortical feed forward projections, which join one cortical area to another primarily through their supragranular layers, and indirect cortico-thalamo-cortical projections, which join infragranular layers of the same original area to the granular layer of the same target area (Fig. \ref{ShermanGuillery}), \citep{Guillery2011}. In Sherman and Guillery's model, direct cortico-cortical projections are ``modulatory,'' providing restricted activation to the target area, and indirect cortico-thalamo-cortical projections are ``driving,'' providing activation across all layers of the target area. Figure \ref{ShermanGuillery} represents Sherman and Guillery's model (based on a simplification of their schematic). We will now describe how this model may be integrated with the Grand Loop.

Recall that each station of the loop in Figure \ref{Loop} is coupled in the feed forward direction. These connections, largely through the supragranular layers, are mirrored in the feedback direction by connections through infragranular layers (Fig. \ref{ShermanGuillery}), \citep{Rempel-Clower2000}. Thus the Grand Loop represents two reciprocal loops, one in the feed forward direction and one in the feedback direction. Furthermore, according to Sherman and Guillery, higher order thalamic nuclei provide at every stage a redundant relay for driving inputs over each feed forward connection. The local cortical circuit then receives signals from these thalamic nuclei and mixes the otherwise independent direct feed forward modulation loop and feedback traversal loop, primarily at layer $4$'s synaptic connections onto supragranular layers, and at supranular layers' onto the infragranular layers' apical dendrites.

We proposed previously that layers $2/3$ of neocortex implement a network for maximizing mutual information between thalamic inputs and cortical responses \citep{Kozloski2007}. Entropy maximization in these layers (equivalent to information maximization when noise in the inputs is assumed to be negligible) would require a dense lateral network \citep{Linsker1997}, which fits well with the high proportion ($\sim22\%$) of total cortical synapses dedicated to intralaminar $2/3$ connections \citep{Binzegger2004}. Given this role for the supragranular layers, we now propose that the role of driving inputs in Sherman and Guillery's model is to provide inputs both from first order thalamic nuclei about the environment and from feedback traversals about the behavioral state of the organism into a global supragranular network that extracts maximally informative features from their combinations. In addition, we propose that these features become conditional modulators on feedback traversals by boosting or reducing the gain on proximal inputs to layer $5$ neurons by means of synaptic inputs onto their apical dendrites from Layer $2/3$ neurons.

\subsection{Basal Ganglia Gating of Feed Forward Functional Pathways}

In Sherman and Guillery's model, thalamic relay neurons in both first order and second order nuclei are subject to modulation. Modulation may derive from direct cortico-thalamic feedback from layer $6$, inhibition from the thalamic reticular nucleus, or from neuromodulatory inputs such as norepinepherine from the locus coeruleus. Sherman and Guillery's model derives largely from their studies of sensory cortices and feed forward pathways through them, projecting from more granular to less granular regions. Here we extend the discussion of thalamic relay neuron modulation to include a role for inhibitory inputs from the basal ganglia to thalamic nuclei that act as relays in the frontal lobe between more granular limbic and motor areas to less granular areas in these regions.

The basal ganglia (including ventral limbic and dorsal motor) are in a privileged position to influence traversals by means of their inhibitory inputs onto thalamic relay neurons in the Grand Loop. These inputs derive from nucleus inominata in the ventral limbic subpallidum and from the globus pallidus in the dorsal motor subpallidum. Neurons in the ventral pallidus (nucleus inominata) receive inhibition from medium spiny neurons (MSNs) in the nucleus accumbens (ventral striatum) and those in the dorsal globus pallidus from those in the dorsal striatum. These neurons then either directly disinhibit thalamic relay neurons or indirectly inhibit thalamic relay neurons through an additional stage of inhibitory neurons (in globus pallidus, this is organized as direct and indirect projections through the external and internal segments). Spiking models of inhibitory pallidothalamic gating have focused on the bird song system \citep{Goldberg2012}, where gating inputs to thalamic relay neurons serve the role of transitioning syllables of the organism's vocalizations. Here we propose a more generic role for this gating in selecting and deselecting different pathways for internal traversals.

Inputs to these direct and indirect pathways through basal ganglia derive from neocortical layer $5$ neurons' projections onto MSNs, and their corticostriatal synapses undergo spike-timing dependendent plasticity (STDP) which is modulated differentially by dopamine depending on the selective expression of either D$1$ dopamine receptors in the direct or D$2$ dopamine receptors in the indirect pathways (Fig. \ref{FDG_Loop}), \citep{Pawlak2008}. Each layer $5$ neuron's collaterals then include a branch descending to the brainstem or spinal cord, a branch descending to thalamus \citep{Guillery2011}, and additionally a branch descending to striatum \citep{Levesque1996}. A recent review of additional types of layer $5$ projection neurons and the role of corticostriatal connectivity in disease provides a thorough examination and schematic of these pathways \citep{Shepherd2013}, and our model of thalamic gating, for now and for simplicity, includes only the ``Pyramidal Tract'' layer $5$ neurons and their projections to basal ganglia and thalamus for the function of thalamic gating.

In summary, our model extends Sherman and Guillery's model of cortico-thalamo-cortical gating of driving, feed forward inputs to include modulation from striatal and pallidal neurons in both the direct and indirect pathways (Fig. \ref{FDG_Loop}). MSNs in our model receive convergent layer $5$ collaterals from all layer $5$ neurons that send convergent collaterals onto a specific thalamic relay neuron. This relay neuron is then gated by the same MSNs, indirectly through globus pallidus (the gate opens for the direct pathway, and closes for the indirect pathway). Such a scheme does not preclude so called ``closed loops'' that originate and terminate in the same cortical area \citep{Kelly2004}, but downplays their significance as only partial regulators of feed forward thalamic gating (Fig. \ref{FDG}A). The basal ganglia in our model is then a ``forward driver gate'' for all feed forward driving signals relayed through the frontal lobe's cortico-thalamo-cortical functional pathways. Because these pathways relay layer $5$ traversals through thalamus to the granular and supragranular layers of cortex, they can indirectly control the routing of feedback signals and the selection of certain traversals over others through the Grand Loop, as we describe in the next section. Additional area to area cortico-thalamo-cortical pathways not on the main loop backbone (such as the visual system) are then available for additional modulation and traversals of the global layer $5$ behavioral network, possibly including loops requiring reafference from the environment.

\subsection{Information Based Gating of Feedback Traversals}

Our model provides two distinct functional signaling pathways through the Grand Loop: feed forward for driving the supragranular entropy maximizing network, and feedback for traversal of the infragranular behavior generation network. The latter, in our implementation of the model, propagates synfire events through a loop, as described by Zheng and Triesch in their model of ``synfire ring'' formation and propagation \citep{Zheng2014}. Restricting synfire activity to the feedback direction is a key aspect of our model. Unlike other models of feedback which ascribe to it solely a sensory processing ``top down'' function, we model the propagation of feedback activity as potentially independent of feed forward activity (for example when a coupling parameter between these two networks is zero). Specifically, the emergence of activations in the supragranular layers are rate coded, while activations in the infragranular layers are spike timing based in order to support synfire events. (We won't speculate here on how these distinct coding schemes are implemented and maintained by the neocortical microcicuit, but it would seem there are ample mechanisms available.)

Conditional coupling between features, extracted by information maximization in the supragranular layers, and spike propagation in the infragranular layers is then under the control of a parameter that models cholinergic modulation in neocortex. Acetylcholine enhances the influence of sensory input on pyramidal cell firing relative to their processing of intrinsic signals within neocortical circuits \citep{Hasselmo2006}. We model this modulatory parameter as changing the slope and dynamic range of the gain function on feedback integration within the synfire ring, such that feature encoding acts as a gate for synfire propagation. As we noted above, this gain function may be implemented by layer $2/3$ inputs to layer $5$ neurons' apical dendrites. The result is that propagation of synfire activity through a column of cortex is \emph{informed} by the categorization of thalamic inputs to that same area. Information maximization among responses in the supragranular areas over environmental inputs becomes entropy maximization of synfire propagation pathways through the infragranular layers, provided that coupling between these is strong (i.e., cholinergic modulation is high). It is because of this coupling that we have named our model an \emph{information based exchange network}.

\subsection{The Forward Driver Gate: Bursting, Modulation, and Plasticity}

Having proposed a central cortico-thalamo-cortical routing function for striatal MSNs by means of their directly disinhibiting or indirectly inhibiting thalamic relay neurons, we will now propose on what basis a striatal MSN adapts to perform this function in the context of system set points. We call this the forward driver gate's ``routing function.'' Our model of MSN firing includes constraints from a weak, assymetric lateral inhibitory network giving rise to ``winnerless competition'' \citep{Rabinovich2001}, and closely matching the periods of striatal bursting lasting hundreds of milliseconds observed \emph{in vivo} \citep{Miller2008}. Ponzi and Wickens have similarly used this network to model spiking properties of striatum \citep{Ponzi2010}, and have shown that at transition points in the lateral network configuration (from low, $\sim10\%$, to high, $\sim20\%$, rates of connectivity), an optimal balance is achieved that facilitates winnerless encoding of variations in driving inputs from neocortex\citep{Ponzi2013}. To achieve this balance, our model instead varies the strength of cortical inputs dynamically by a dual source of modulation of STDP at the corticostriatal synapse.

The first dynamic modulator of STDP at the corticostriatal synapse in our model is GABA inhibition from the lateral network, itself responsible for ``turn-taking'' among MSNs and their bursts, characteristic of the winnerless network. We assume that both direct and indirect pathways show STPD reversal under GABA inhibition \citep{Fino2010, Paille2013}, and we model winnerless competition between striatal neurons as the source for this inhibition (Fig. \ref{FDG}B).

The second dynamic modulator of STDP at the corticostriatal synapse is dopamine. Given the routing function's potential as a critical determiner of the emergence of behavior, affect, and cognition in the organism via its direct control over traversals of the layer $5$ network, reward-based learning of this function is ultimately required. For now, we simulate our brain model of information based exchange with dopamine-based learning serving only a closed-loop function, separate from the environment and therefore independent of reward encoding. This closed-loop function is sensitive to system set points and monitors traversals. It is equivalent to so-called ``tonic firing'' in dopamine neurons, which can also include bursts. We propose that the intrinsic dynamics of dopamine neuron membrane currents implements this closed-loop function by measuring time and the abruptness of changes to system states, with bursts generated under specific conditions summarized below. Dopamine provides a potent modulation of STDP at the corticostriatal synapse \citep{Pawlak2008}, and further modulates it differentially at the inputs to D$1$-MSNs and D$2$-MSNs. In our model this differential modulation, combined with GABA modulation, produces the complex routing function summarized in Figure \ref{FDG}C.

To signal basal reward inputs to the organism, dopamine neurons have been shown to fire bursts of action potentials in response to strong excitatory inputs to medial tegmentum and substantia nigra pars compacta (SNc). Because of these responses, the dopamine system has been extensively modeled as recapitulating reinforcement learning and operant conditioning in the organism. We propose here for the first time an additional closed-loop role for dopamine neurons in learning routing functions and selecting traversals. Specifially, we propose that dopamine neurons signal changes to traversals, and thereby influence the subsequent emergence of new traversals. The basis for this proposal derives from recent connectomics studies, which demonstrate that fully $70\%$ of dopamine neuron inputs are inhibitory, and that most of this inhibition arises from striatum \citep{Watabe-Uchida2012}. Our closed-loop role for dopamine modulation depends on this inhibition, and this proportion and source suggests that closed-loop responses to inhibitory inputs, not open loop responses to basal excitation and reward, may be the predominant operating mode of the dopaminergic system.

Dopamine neurons exhibit heterogeneous combinations of intrinsic $\mathrm{I{_h}}$ and $\mathrm{I{_A}}$ currents \citep{Amendola2012}, as well as T-type calcium currents, which together generate post inhibitory rebound bursting in slice preparations. These currents' role \emph{in vivo} has not yet been demonstrated, but our model assumes that the dynamical criteria for dopamine neuron bursting (and subsequent learning of routing at corticostriatal synapses) are implemented by rebound bursting. Other models have explored rebound bursting in dopamine neurons \citep{Lobb2011}, but not in the context of a closed-loop regulatory function. In our model, if the duration and abruptness of removal of striatal inhibition to dopamine neurons is appropriate, a rebound burst occurs. This aspect of the model indirectly imposes the additional criterion that inputs from layer $5$ to striatum that transition the MSN winnerless network should be similarly matched to the duration and abruptness of change required for rebound spiking in dopamine neurons. In this way the striato-nigro-striatal loop monitors changes in traversals and alters routing within the feed forward entropy maximizing network by modulating corticostriatal STDP. We therefore propose that MSNs learn this routing based in part on their ability to recognize patterns of spiking in layer $5$ that remain stable for a minimum duration of time then fade rapidly, a property expected during traversals of the Grand Loop.

\section{Simulation Methods}

We simulated the model to explore its dynamics, characterize preliminary set points for measurement and analysis, and study traversal behavior and it's regulation under different modulatory conditions. The five major components of the model to be simulated included cortical layers $2/3$, $5$, thalamus, striatum, and dopamine neurons. Meeting this challenge at the detailed level of neural tissue simulation is beyond the scope of this report, and without a good understanding of target model set points, likely impossible. We therefore aimed to draw upon four simplified abstractions of the key behaviors we ascribe to principle cells in these structures \citep{Linsker1997,Zheng2014,Rabinovich2001,Mihalas2009}. With four base component models replicated from other studies, we then coupled them across novel interfaces, realizing the closed, functioning Grand Loop, complete with its subcortical regulators.

\begin{table}[!t]
\textbf{\refstepcounter{table}\label{Params} Table \arabic{table}.}{ Component Models and Parameters }

\processtable{ }
{\begin{tabular}{lllc}\toprule
Component & References & Parameter Name & Parameter Value \\\midrule
Neocortex, layers $4$ and $2/3$ & \cite{Bell1995};& $\beta_{x_0}$ & $0.00002$ \\
& \cite{Linsker1997};& $\beta_{y_0}$ & $0.0007$ \\
& \cite{Kozloski2007}& $\beta_C$ & $0.0007$ \\\midrule
Neocortex, layer $5$ & \cite{Zheng2014}& $\eta_{\mathrm{IP}}$ & $0.01$ \\
& & ${T^E}_{\mathrm{max}}$ & $1.0$ \\
& & ${T^I}_{max}$ & $0.5$ \\
& & $\mu_{\mathrm{IP}}$ & $0.1$ \\
& & $\sigma_{\mathrm{HIP}}$ & $0$ \\
& & $\eta_{\mathrm{inhib}}$ & $0.001$ \\
& & $\sigma^2_\xi$ & $0.01$ \\
& & $\eta_{\mathrm{STDP}}$ & $0.004$ \\
& & $\eta_{\mathrm{iSTDP}}$ & $11.0$ \\\midrule
Striatum & \cite{Rabinovich2001}& $g_{\mathrm{max}}$ & $0.25$ \\
& & $g_{\mathrm{min}}$ & $0$ \\
& & $a$ & $0.7$ \\
& & $b$ & $0.8$ \\
& & $\tau_{1}$ & $0.08$ \\
& & $\tau_{2}$ & $4.1$ \\
& & $\nu$ & $-1.5$ \\
& & $x_{0}$ & $-1.2$ \\
& & $y_{0}$ & $-1.62$ \\
& & $z_{0}$ & $0$ \\\midrule
Dopamine Neurons & \cite{Mihalas2009}& $b$ & $1.0$ \\
& & $G/C$ & $50$ \\
& & $k_{1}$ & $200$ \\
& & $k_{2}$ & $20$ \\
& & $\Theta_{\inf}$ & $-0.05$ \\
& & $R_{1}$ & $0$ \\
& & $R_{2}$ & $1.0$ \\
& & $E_L$ & $-0.07$ \\
& & $V_R$ & $-0.07$ \\
& & $\Theta_r$ & $-0.06$ \\
& & $a$ & $1.0$ \\
& & $A_{1}$ & $5.0$ \\
& & $A_{2}$ & $-0.3$ \\\botrule
\end{tabular}}{}
\end{table}

\subsection{Component Models}

Four component models from the literature were targeted here to capture the functions of cortical layers $2/3$ and $5$, striatum, and dopamine neurons in the brain model. These four met sufficient requirements to implement information based exchange, with very few changes to published parameters. We list the models below and describe the requirements they satisfy. Parameters defined in the original references for each component model are found in Table \ref{Params}. Because thalamic relay neurons were implemented as a simple set of sums over inputs, they are described as an interface between component models in the subsequent section.

\begin{itemize}
\item \textbf{Neocortex, layer $\textbf{2/3}$}: The model applies the ``Infomax'' algorithm \citep{Bell1995} to thalamic relay neuron inputs. A neural network implementation of the same optimization \citep{Linsker1997} based entirely on a local learning rule, establishes the biological plausibility of this function \citep{Kozloski2007}. In brief, the algorithm takes the full rank weight matrix $C$ and inverts its transpose to compute a Hebbian learning rule with the term $({C^T})^{-1}$ for entropy maximization over an ensemble of input vectors $x\in X$ to a neural network. The input vector $x$ is offset by $x_0$ to have zero mean. The output vector of the network $Cx$ is transformed by the learned offset $y_0$ and a nonlinear logistic function to becomes the layer $2/3$ area's output $y \in (0,1)$, which maximizes the mutual information over the input ensemble. The input offset, the offset of the neural network output, and the weight matrix, are each updated with learning rates $\beta_{x_0}$, $\beta_{y_0}$, and $\beta_C$ (Table \ref{Params}).
\item \textbf{Neocortex, layer $\textbf{5}$}: The model evolves from a self-organizing recurrent network of binary spiking units through application of homeostatic plasticity, weight normalization, and STDP learning rules, together with synaptic pruning and synaptogenesis \citep{Zheng2014}. This biologically consistent set of synaptic modifications creates distributions of synaptic densities and weights that evolve over time to closely match data from
    developing neocortex. The weight matrix $W$ also develops robust feed forward motifs and synfire activity similar to the model of \cite{Kozloski2010}, but with the remarkable topological feature of a closed, global loop of distinct propagation layers (Fig. \ref{L5}), which together engender ``synfire rings.'' We evolved this network for $200,000$ time steps ($\Delta t=1$ msec) to create four areas of cortex, which were then embedded into the larger model as two frontal lobe ($M_1$, $M_{\mathrm{sup}}$) and two sensory lobe ($S_1$, $S_{\mathrm{sec}}$) areas. Weights close to zero were held at zero for the remainder of all simulations. Propagating activity is maintained in the excitatory network, satisfying the requirement for layer $5$ traversals. An inhibitory network that undergoes biologically plausible inhibitory STDP at its synapses onto excitatory neurons, together with homeostatic plasticity in the excitatory network, maintains activity in the synfire ring at a nominal firing rate of $100$ spikes/sec. The inhibitory network imposes global, persistent competition across the network of excitatory layers. We propose this inhibition as an approximate functional model of inhibition from the thalamic reticular nucleus, which also integrates activity from across the thalamocortical system.
\item \textbf{Striatum}: The model creates activation paths within the state space of a weakly connected, asymmetric inhibitory network to give rise to ``winnerless competition'' \citep{Rabinovich2001}, and alternating bouts of activity among the different neurons in the network (Fig. \ref{StrBursts}). These bouts have been used by others to model the intrinsic dynamics of the striatum \citep{Ponzi2010}, and together represent global attractor states that encode the modulatory and reorganizing influence of excitatory inputs to the network from cortical layer $5$. Using a FitzHugh-Nagumo model, MSNs are represented by three dynamic variables. First, $x_f(t)$ in the model represents the ``burst potential'' of the neuron, with a positive transient in this potential representing a $\sim 350$ msec burst. Computed using the same time step as the binary spiking layer $5$ model, this coarse resolution model of the neuron's membrane potential is appropriate given the dominant bursting mode of firing in MSNs, and the observation that activity is often observed as alternating series of bursts of bursts \citep{Miller2008}. The remaining variables $y_f(t)$ represent a recovery from inhibition and $z_f(t)$ the inhibitory synaptic current received by the neuron, summed over the inhibitory inputs from other neurons through a heaviside step function and the inhibitory weights $g$.
\item \textbf{Dopamine Neurons}: The model is that of a leaky integrate and fire neuron. Four state variables are computed: a membrane potential $V(t)$, a variable threshold $\Theta(t)$, and two intrinsic currents $I_1(t)$ and $I_2(t)$, each integrated over the same time step as the previous two models. Because spikes in this model are represented by instantaneous resets of each variable at $V(t)>\Theta(t)$, the time step ($\Delta t=1$ msec) is sufficient to integrate the neuron's spiking dynamics. Based on the published model, we derived an instance of a ``rebound burst'' model, and satisfied the requirements of dopamine neurons in the closed striato-nigro-striatal loop. Specifically, the voltage-dependence of $\Theta(t)$ permits the model to generate rebound spiking under conditions when the neuron has been hyperpolarized deeply, or for a prolonged period (Fig. \ref{DA}). Due to the independent spike-induced current $R_2$, each rebound event generates a burst of four action potentials. This simplification's phenomenology also approximates that generated by other more complex models of rebound firing in dopamine neurons \citep{Lobb2011}.
\end{itemize}

\subsection{Component Model Interfaces}

The interfaces between component models that create the integrated brain model of information based exchange are now listed and described.

\begin{itemize}
\item \textbf{Feedback Cortico-cortical}: Layer $5$ feedback inputs to a cortical area layer $5$ are modeled as in the self organizing recurrent network of \cite{Zheng2014} to implement the traversal network. Inputs are also categorized by the layer $2/3$ model as an input vector of sums of binary spike trains over a time window $\tau_X$. This vector $\hat{x}_{\mathrm{FB}}(t)$ comprises the elements $\hat{x}_{{\mathrm{FB}}_i}(t) \leftarrow \sum_{t-\tau_X}^{t} s_{{\mathrm{FB}}_i}(T)$, where $s_{{\mathrm{FB}}_i}(T)\in \{0,1\}$ is the spike train from one unit in the upstream layer $5$ area. Because of the full rank requirement of the information maximizing algorithm, the model of layer $2/3$ includes a fixed first stage random mixing matrix $M_{\mathrm{FB}}$, drawn from a lognormal distribution with unit mean and unit standard deviation, which linearly combines the elements of $\hat{x}_{\mathrm{FB}}$ to create the feedback input vector $x_{\mathrm{FB}}(t) \leftarrow M_{\mathrm{FB}}\cdot \hat{x}_{\mathrm{FB}}(t)$.
\item \textbf{Feed Forward Cortico-thalamo-cortical}: Inputs to a thalamic relay neuron $j$ projecting to a cortical area are modeled as a vector of sums over a time window $\tau_X$ of binary spike trains from layer $5$ units in the cortical area projecting in the feedforward direction to the same area. This vector $\hat{x}_{\mathrm{FF}}(t)$ then comprises elements $\hat{x}_{{\mathrm{FF}}_i}(t) \leftarrow \sum_{t-\tau_X}^{t} s_{{\mathrm{FF}}_i}(T)$, and is similarly transformed by a mixing matrix such that the thalamic relay neuron's activity $\theta_{\mathrm{FF}}(t) \leftarrow M_{\mathrm{FF}} \cdot \hat{x}_{\mathrm{FF}}(t)$. Each element $\theta_{\mathrm{FF}_j}(t)$ is then subjected to the forward driver gating vector $G$, such that elements of the feed forward input vector are $x_{\mathrm{FF}_j}(t) \leftarrow G_j \cdot \theta_{\mathrm{FF}_j}(t)$.
\item \textbf{Layer $\textbf{4}$ Thalamic and Feedback Inputs} : Feed forward thalamic inputs to layer $4$ are combined with feedback inputs, such that the input vector to information maximization in layer $2/3$, $x \leftarrow x_{\mathrm{FB}}+x_{\mathrm{FF}}$. It is at this stage also that sensory inputs from a simulated environment may be added to the model.
\item \textbf{Layer $\textbf{2/3}$ to Layer $\textbf{5}$}: The Layer $2/3$ output vector $y$ provides an input to a gain function for layer $5$'s integration of binary spikes from feedback traversals of the Grand Loop. This gain function is a model of the layer $5$ neuron's apical dendrite, and is parameterized by the term ${Ach} \in [0,1]$, a proxy for the level of cholinergic modulation in neocortex. The gain on inputs to layer $5$ unit $j$ is then $U_j=[1-Ach(1-y_j)]/[1-Ach/2]$, which at $Ach=0$, preserves unitary gain regardless of $y$, and at $Ach=1$ provides a gain $U\in(0,2)$ for $y \in (0,1)$. In this way, assuming information maximization divides the population into different halves of active and inactive units, the total synaptic input to the network will remain constant, since the $U_j$ will always have a mean of $1$, and is applied multiplicatively to the excitatory synaptic integration function of each layer $5$ neuron as in \citep{Zheng2014}.
\item \textbf{Globus Pallidus to Thalamus}: The forward driver gating function $G \in \{0,1\}$ applied to thalamic relay neurons in the feed forward cortico-thalamo-cortical pathway models the final output of basal ganglia, a transient increase (via the indirect pathway) or decrease (via the direct pathway) in inhibition. $G$ is computed using a modified pallido-thalamic adjacency matrix $D$, comprising $1$ for all direct pathway pallido-thalamic inputs, $-1$ for all indirect pathway pallido-thalamic inputs, and $0$ for all unconnected pallidal to thalamic relay neurons. The bursting outputs of MSNs are represented by the half wave rectification function $V$, of the burst potential variable $x_f(t)$, and the gating function is then written $G=\mathcal{H}[D\cdot V(x_f(t))]$, where $\mathcal{H}$ is the heaviside function.
\item \textbf{Layer $\textbf{5}$ to Striatum}: The inputs from the Layer $5$ model to an MSN in the Striatum model are drawn from all layer $5$ neurons in the cortical area for which the MSN gates inputs at the thalamus, and from those in the areas connected to it in either the feed forward or feedback directions. These Layer $5$ inputs may also be directed to motor outputs of the model to a simulated environment (as in the Pyramidal tract). Corticostriatal synapses are subjected to STDP that differentially adjusts weights based on correlation between cortical spiking and the derivative of the burst outputs of MSNs: $V^{\prime}(x_f(t))$. Pre-post pairing is defined as when a cortical spike occurs and this derivative is positive, and post-pre pairing when a cortical spike occurs and it is negative. Each kind of paring is computed separately and subjected to the modulatory conditions at the synapse, as illustrated in Fig. \ref{FDG}. Briefly, depending on $1$. the identity of the MSN (D$1$- or D$2$-type), $2$. whether dopamine is or is not present at the synapse, and $3$. whether the inhibitory synaptic current $z_f(t)$ at the MSN exceeds a threshold ($z_f(t)>0.00707$), each pairing value may be either $1$ or $0$, and the adjustment to the weight a multiple of this value and a learning rate of 0.002. As in \cite{Zheng2014}, weights are normalized such that the sum of all inputs to an MSN cannot exceed $0.1$. When weights reach zero they are pruned, and new connections may then be formed during a time step with probability $0.2$.
\item \textbf{Striatum to Dopamine Neurons}: The input to each dopamine neuron in the model, $I_e$ \citep{Mihalas2009}, is computed by summing all burst potentials from those MSNs projecting to the dopamine neuron, multiplied by a constant weight of $-2.25$.
\item \textbf{Dopamine Release to Corticostriatal Synapses}: Unlike all other projections in the model's interfaces, the dopamine neuron projection is to a synapse, not a neuron. Specifically, dopamine spiking results in a persistent dopamine modulation of STDP at a specific set of corticostriatal synapses. Dopamine neurons are assigned randomly without replacement to corticostriatal synapses onto each MSN. The duration of dopamine modulation following a Dopamine Neuron model spike persists at the synapse for a time $\tau_{\mathrm{DA}}$.
\end{itemize}

\subsection{Simulated Experiments}

We simulated the model to explore the rate and heterogeneity of transitions in traversals and in subcortical modulators of these traversals. The configuration (Table \ref{Configuration}) allowed for a rapid prototyping because of the simulation's small size. Following the initialization of the cortico-cortical Grand Loop network of four areas, we simulated the larger model for an additional $500,000$ iterations. The first $50,000$ iterations were used to adjust the biases of the layer $2/3$ model, during which time $Ach$ modulation of layer $5$ was drawn from the positive half of a zero mean normal distribution with standard deviation of $1$. All plots, except where noted, show the final iterations of the $500,000$ total. Reported are experiments wherein the parameter $\tau_{\mathrm{DA}}$ was set at either $25$ or $100$ msec, and $Ach$ at $0.25$, $0.5$, or $0.75$. All plots except Figure \ref{Ach} show results for $\tau_{\mathrm{DA}}=100$ msec.

\begin{table}[!t]
\textbf{\refstepcounter{table}\label{Configuration} Table \arabic{table}.}{ Configuration Parameters Specific to Model Simulation}

\processtable{}
{\begin{tabular}{llc}\toprule
Parameter Name & Description & Parameter Value \\\midrule
 $N_A$ & Number of cortical areas & $4$ \\
 $N_E$ & Number of thalamocortical units (layer $5$, $2/3$ pyramidal, thalamic relay neurons) & $400$ \\
 $\tau_X$ & Layer $2/3$ integration window for spiking to rate code transformation (msec) & 100 \\
 $N_I$ & Number of thalamic reticular inhibitory neurons & $80$ \\
 $N_{\mathrm{Frontal}}$ & Number of frontal cortical areas under striatal gating & $2$ \\
 $N_{\mathrm{Str}}$ & Number of striatal MSNs & $100$ \\
 $N_{\mathrm{Cx,Str}}$ & Number of cortical neurons projecting to a striatal neuron  & $20$ \\
 $W_{0_{\mathrm{Cx,Str}}}$ & Initial corticostriatal weight & $0.005$, ($0.1/N_{\mathrm{Cx,Str}}$) \\
 $N_{\mathrm{Str,Th}}$ & Number of striatal neurons projecting to a thalamic relay neuron & $11$ \\
 $N_{\mathrm{DA}}$ & Number of dopamine neurons & $20$ \\
 $N_{\mathrm{Str,DA}}$ & Number of striatal neurons projecting to a dopamine neuron  & $20$ \\
 $\tau_{\mathrm{DA}}$ & Dopamine modulation window (msec) & $25, 100$
\\\botrule
\end{tabular}}{}
\end{table}

\section{Simulation Results}

\subsection{Coordinated Behavior Among Component Models}
Behavior of the model may be analyzed first based on inspection of various raster plots from different components of the model. In this way coordination between the different components is apparent. We first observed that cortico-cortical traversals through the feedback layer $5$ network occur without subcortical regulation, and were similar to the synfire events reported by \cite{Zheng2014}. There are two main regulators of these traversals in our model: $1$. an information based gain on layer $5$ feedback inputs provided from layer $2/3$, and $2$. basal ganglia gating of cortico-thalamo-cortical feed forward inputs to layer $2/3$ information maximization by the forward driver gate.

Upon introducing these regulators, we noted that traversals became structured into long bouts of smoothly alternating and repeating patterns of activity across the different cortical layers' raster plots. Each pattern persisted for $\sim 400$ msec (Fig. \ref{Raster}A), and sequences of patterns, while similar over each cycle, were not identical. The $Ach$ parameter provides a means to adjust the influence of categories learned by layer $2/3$ on traversals. For this initial experiment, $Ach=0.25$ provided a gain $U \in (0.86,1.14)$ for $y \in (0,1)$.

Information maximization creates maximal entropy in the ensemble of output of vectors over an input ensemble, and because of the logistic function, activity in each layer $2/3$ neuron was typically close to zero or one. We interpret these values as cortical up and down states, which have both an extrinsic and intrinsic origin in the local cortical microcircuit.

Maximizing entropy of the ensemble of gain functions in this way, applied to layer $5$ inputs in the feedback traversal network, had the interesting effect of creating more irregularity in the patterns of activity across all of cortex as $Ach$ increased. At $Ach=0.5$, $U \in (0.67, 1.33)$, (Fig. \ref{Raster}B), pattern combinations became varied, even though average global firing rates imposed by homeostatic plasticity in the network were consistently maintained ($100$ spiked/sec). Finally, at $Ach=0.75$, $U \in (0.4, 1.6)$, traversal transition rates increase significantly, and patterns were highly varied (Fig. \ref{Raster}C).

Inspecting the information bearing up and down states in layer $2/3$ directly in state rasters from all four cortical areas also reveals coordination between areas and with transitions in traversals. In Figure \ref{Up}A., under $Ach=0.25$, the rate of state changes among layer $2/3$ units appeared coordinated, especially in the secondary sensory area. This coordination is less regularly transitioned than in the traversals, and occurs at a higher rate. At higher $Ach=0.5$ (Fig. \ref{Up}B) up and down state coordination with traversals increases, while coordination across layer $2/3$ is weakened. At $Ach=0.75$ (Fig. \ref{Up}C) states becomes synchronized in the secondary sensory area and more coordinated with traversals overall, even though traversals themselves become more heterogeneous. Note that the heterogeneity in traversals due to increased control by the information maximizing network is not due to a lack of convergence in the weights of the networks. Weights among both the layer $2/3$ Infomax input weights $C$ and layer $5$ feedback weights $W$ converged during these simulations.

MSN bursts generated by the model were ongoing, as in the winnerless network and the model of \cite{Ponzi2010}. These bursts appeared in fast sequences, which were of longer duration in D$1$-type MSNs than D$2$-type (Fig. \ref{StrRaster}). Variability in burst rate between MSNs was also observed, with some not firing at all, likely because of inhibition from the active network. Increasing $Ach$ had only a small effect on the raster appearance, and so we began our quantitative analysis by examining coordination between the Striatum model and the Layer $5$ model.

\subsection{Measurements of Information Based Exchange}
To quantify coordination between striatum bursting and cortical layer $5$ spiking, we computed pairwise linear correlation coefficients between each cortical spike train and striatal burst train. We plotted each using a color scale (red, more correlated; blue, less correlated) in a matrix showing how different areas of cortex fired in relation to D$1$- and D$2$-type MSN bursts (Fig. \ref{Corr}, left column). Only significant correlations were plotted, and all others were represented by zero. We also show that the mean of each distribution of correlation values (Fig \ref{Corr}, right column) for both D$1$- (blue) and D$2$-type (red) MSNs differed. Most coefficient distributions of D$1$ vs. D$2$ burst correlation with cortical spiking were significantly different($p<0.05$), based on pairwise student t-tests. More striking is the difference in sign for each mean coefficient of correlation to each cortical area as $Ach$ increases. Positive correlation coefficients dominated at low $Ach$ and negative at high. At the intermediate level, M$1$ in particular showed a divergence in sign between mean correlation coefficients for D$1$-type (positive) and D$2$-type (negative) MSNs.

Finally, to quantify information based exchange directly, we measured the entropy of cortical spiking and dopamine neuron spiking, and the mutual information between cortical and dopamine neuron spiking \citep{Strong1998}. Instead of measuring entropy and information among spike trains of individual neurons however, which quantify the distribution of patterns of spikes over time, we measured entropy and information in population spiking, which quantify the distribution of patterns of spikes over the population for single time steps. The method was aimed at asking if traversals themselves show entropy maximization based on increased modulation from layer $2/3$. Synfire events are encoded by the sets of units that participate at every stage of the chain or ring propagation. Therefore, if the entropy of synfire population spiking increases, it can be concluded that the synfire chain entropy itself has increased.

We found that entropy in cortical layer $5$ population spiking increased as $Ach$ increased (Fig. \ref{Ach}A). We also show that as the window of dopamine integration $\tau_{\mathrm{DA}}$ increased, the entropy of layer $5$ population spiking increased slightly as well. Surprisingly, the entropy of dopamine neuron population spiking (Fig. \ref{Ach}B) remained constant while both parameters in the model were altered. Finally, to measure how increasing traversal entropy depends on dopamine population spiking, we measured the mutual information between these two populations, and found it to decrease as $Ach$ increased (Fig. \ref{Ach}C).

\section{Discussion}

We discuss the brain model of information based exchange in three contexts: brain evolution and development, brain resting state networks, and new approaches to the study of brain disorders such as neurodegenerative diseases.

\subsection{Brain Evolution and Development}

We propose that the Grand Loop, spanning sensory, limbic, and motor cortices, and specifically traversing in our model \emph{somatosensory} cortices, is prototypical and embryonic in origin, since other modalities develop fully only after birth and do not share a granular-agranular tiling boundary in von Economo's map. The topological relationship between other modalities and this backbone may then provide alternative pathways for completing a full traversal and rapidly binding percepts, needs, and behaviors. Finally, the tight coupling between somatosensory inputs and limbic states (i.e., tissue damage, pain) and motor states (i.e., sensorimotor feedback, proprioception) argues that this loop is likely preeminent in both brain evolution, organization, and development.

This model additionally provides insights into those organisms lacking cortices, wherein the stages of the proposed traversals may not be segregated anatomically (e.g., into Brodmann areas), but instead may be nucleated (e.g., in the birdsong system), or even superimposed within the same pallidal regions (e.g., in fish and amphibians). Synfire ring development is robust given the synaptic modifications proposed by \cite{Zheng2014}. It furthermore does not require anatomical segregation between layers to emerge, nor for synfire activity to propagate (e.g., for Fig. \ref{L5}, we sorted each matrix after areas developed in order to illustrate them clearly and connect subcortical structures to each).

Synfire rings may represent a prototypical substrate for behavior generation (Fig. \ref{Map}), and through subpallidal regulatory inputs from thalamus and basal ganglia as described herein, for behavior selection. In such a scenerio, the evolution of a multilaminate neocortex to support such rings may have solved the problem of entropy maximization over the ensemble of synfire events in very large networks. Since the neural network implementation of Infomax requires a dense lateral network, to optimize each stage of a synfire ring and traversals in general would necessarily require segregation of stages and a superimposed information maximizing network (Fig. \ref{Map}). This solution to the problem would support rapid expansion of the synfire ring substrate by evolution, given that redundancy in large networks could suddenly be managed and eliminated by information maximization.

\subsection{Resting State Networks}

The challenge of modeling resting state activity in the brain has presented itself based on observations that distinct networks spanning multiple cortical areas appear in imaging studies to serve either active or inactive states of the organism \citep{Fox2005}. Inactivity correlated networks appear even under anesthesia \citep{Vincent2007}, and these areas have very high metabolic rates, tipping the brain's energy budget towards a large investment in the organism doing nothing.

What this costly outlay accomplishes may be explained by our model's use of closed-loop activity in the information based exchange network to increase entropy over the ensemble of traversals. In an evolutionary context, this activity may be viewed as preadapting the brain to selecting novel behaviors in novel contexts by maximizing such a quantity first, before engaging with the environment, then using the diverse traversals to explore it and seek reward.

While others have noted that resting state dynamics may represent a ``constant state of inner exploration'' \citep{Deco2011}, our model is the first to assign a quantitative measure to the fruits of this brain activity, providing a new way to reason about the trade off between evolutionary pressures towards latent adaptive behaviors and the large metabolic cost of resting state network activity.

\subsection{Dynamic Disease Risk}

We hypothesize that basic controls are required to establish ``cognitive homeostasis,'' i.e., a process by which variables that change brain dynamics are carefully regulated so that properties of brain state transitions (and thus brain information processing and behavioral dynamics) remain relatively stable under constant neuromodulatory conditions. We refer to these stable properties as ``set points,'' i.e., targeted norms for critical system variables supporting normal behavior, percepts, affect, and cognition. In our model, these controls are based on a consistent set of parameters that yield consistent spiking and bursting patterns, even when the network undergoes reorganization (e.g., when $Ach$ was modified, the system adjusted and produced stable traversals). Stable ranges of firing among burst rates and traversals, coefficients of correlated firing and bursts, and entropy and mutual information among population spiking and bursting have been our initial targets for describing these system set points using the brain model.

In real brains, given evolutionary pressure for robust self-regulation and behavior, the system is certainly replete with controls aimed at maintaining these set points. The challenge of studying brain disorders such as neurodegenerative disease is sorting primary and secondary risks from the multitude of compensatory mechanisms, each of which manifests itself as a deviation from normal brain and neuronal function given some primary genetic or injury risk. Researchers have shown, for example, that mutant Huntingtin protein disturbs NMDA receptor localization, densities, and currents at the corticostriatal synapse in mouse models of the disease \citep{Cepeda2001}. Knowing how this change arises and perturbs circuit dynamics, plasticity, and system set points may provide a better understanding of why certain neurons succumb and others don't when subjected to the same mutant protein.

We propose that perturbations in our model may result in stable dynamics, but with measurable risks related to stressors on normal neuronal function. If these risks are extreme in our model, and therefore difficult to compensate for in biological tissue, a cascade of neuronal dysfunction may result. Neurodegenerative diseases such as Huntington's, Parkinson's, and Alzheimer's, may then be understood as cascading failures given initial stressors derived from plasticity abnormalities at the corticostriatal synapse, in the striato-nigro-striatal loop, and over the process of entropy maximization in layer $2/3$, respectively. For example, subtle changes to STDP or homeostatic plasticity may result in increased synaptic competition or cycling in the space of possible weights, which is then difficult to compensate for locally, given that traversals entail global brain states. If these risks increase when stressed neurons are removed from a simulation, the model may then be used to predict disease progression.

Implementation of the current brain model of information based exchange forms a framework for the analysis of cognitive homeostasis in disease using IBM's scalable approach to structural and neurophysiological modeling of neocortex and brain nuclei \citep{Kozloski2011}. Here we extend this approach and that of many brain modeling projects, which seem focused on validating complex local circuit and tissue models at the expense of validating tissue inputs. Minimal complexity brain models, in our case an information based exchange network, may be necessary to capture brain dynamics and provide validatable inputs to complex tissue models. With this new approach, inputs and models of the various components may be validated against \emph{in vivo} experimental observations and simulated over very long time scales in order to stress the model and its set points in physiologically and clinically realistic ways.

Additional perturbations to the model may include physiological stimulation, such as simulated deep brain stimulation (DBS) in simulated neural tissue, drugs with known targets in the detailed model, and different disease states with hypothesized mechanisms at the level of gene, protein, regulatory network, etc. Stimulation, drug effects, and disease mechanisms can then be targeted to test certain hypotheses about modifications to dynamic disease risk, and to study the wider system's behavior. Increasing complexity of perturbation sets (targets and combinations) may be designed to validate the model under different therapeutic conditions, and to test for phenotypic outcomes (e.g., symptomatology).

In the above discussion, a model of several brain circuit components and their global set points is proposed as a means to test disease mechanisms and therapeutic inputs such as DBS and drugs. The implicit assumption of these tests is that risks can be inferred from outlier variables that maintain system set points, and that these outliers may then be implicated as causes of phenotypic symptoms such as abnormal behavior at the organismal, circuit, neuronal, or synapse level. Targeting these variables in real world systems is one approach we propose for novel therapeutic design and discovery using brain modeling combined with neural tissue simulation.

\section*{Disclosure/Conflict-of-Interest Statement}

The authors declare that the research was conducted in the absence of any commercial or financial relationships that could be construed as a potential conflict of interest.

\section*{Acknowledgments}
We acknowledge many years of helpful discussions with Charles C. Peck and Guillermo Cecchi. We also acknowledge discussions with Ralph Linsker and Roger Traub on component model design, and Robert Rogers and Robert Kerr on the integrated model design. We acknowledge helpful comments on the manuscript from Erik Schomburg, Tuan Hoang Trong, and Pengsheng Zheng. Finally, we acknowledge the artistic contributions of Stella Kozloski, who produced the drawings based on von Economo's work.

\textit{Funding\textcolon} A portion of this work was funded by CHDI.

%

\bibliographystyle{plainnat}
\bibliography{ibex}


\section*{Figures}


\begin{figure}[h!]
\begin{center}
\includegraphics[width=18cm]{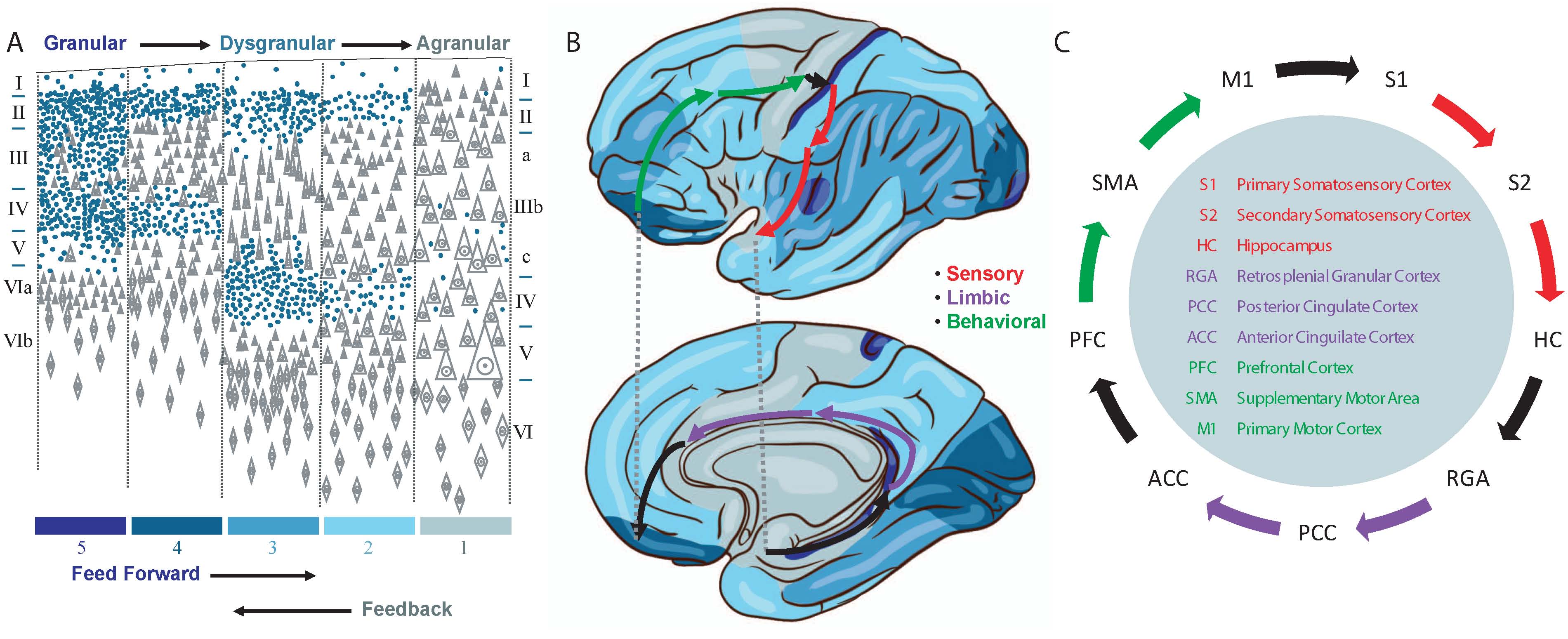}
\end{center}
\textbf{\refstepcounter{figure}\label{Loop} Figure \arabic{figure}.}{ A. Granularity of different neorcortical areas, adapted from \cite{vonEconomo}. Colors at bottom correspond to the map in B. B. von Economo's neocortical tiling based on the granularity of large regions of neocortex spanning multiple Brodmann areas. The location of three Brodmann areas per stage are waypoints along a feed forward Grand Loop (arrows). C. These Brodmann areas are connected based on projection data. Evidence that feed forward connections progress from granular to agranular areas provides directionality. The reciprocal feedback loop is not shown.}
\\\\
\begin{center}
\includegraphics[width=19cm]{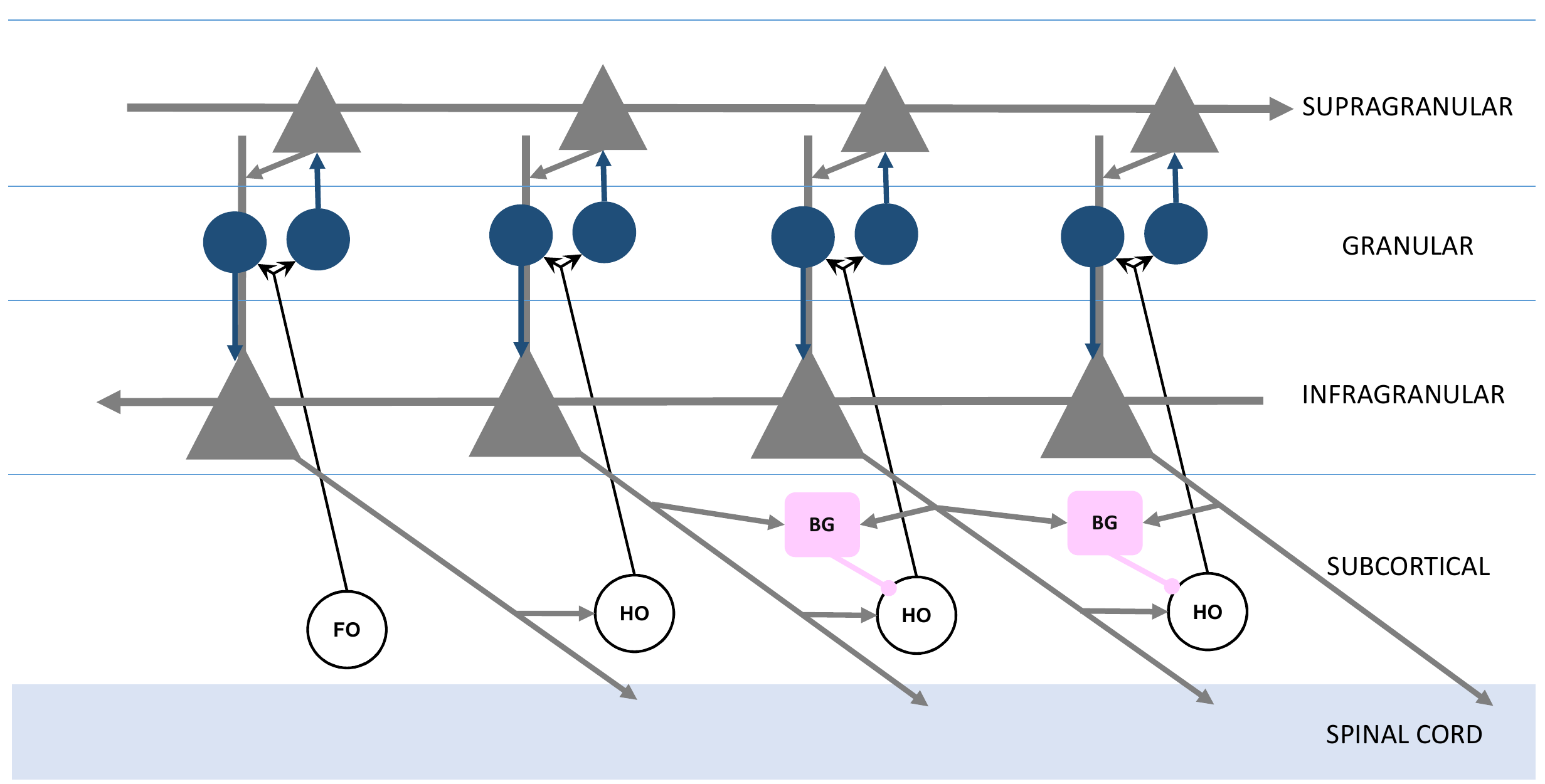}
\end{center}
\textbf{\refstepcounter{figure}\label{ShermanGuillery} Figure \arabic{figure}.}{ Organization of feedforward and feed back functional connections, adapted from \cite {Guillery2011}. Infragranular and supragranular layer pyramidal neurons (gray triangles) form direct feedback and feed forward connections, with the local circuitry receiving first order (FO) and higher order (HO) thalamic nuclei inputs through granular layer spiny stellate neurons (blue circles). The basal ganglia (BG, pink boxes) receive infragranular inputs, and provide inhibitory gating to higher order nuclei in the brain's frontal lobe. }
\end{figure}
\newpage
\begin{figure}[h!]
\begin{center}
\includegraphics[width=19cm]{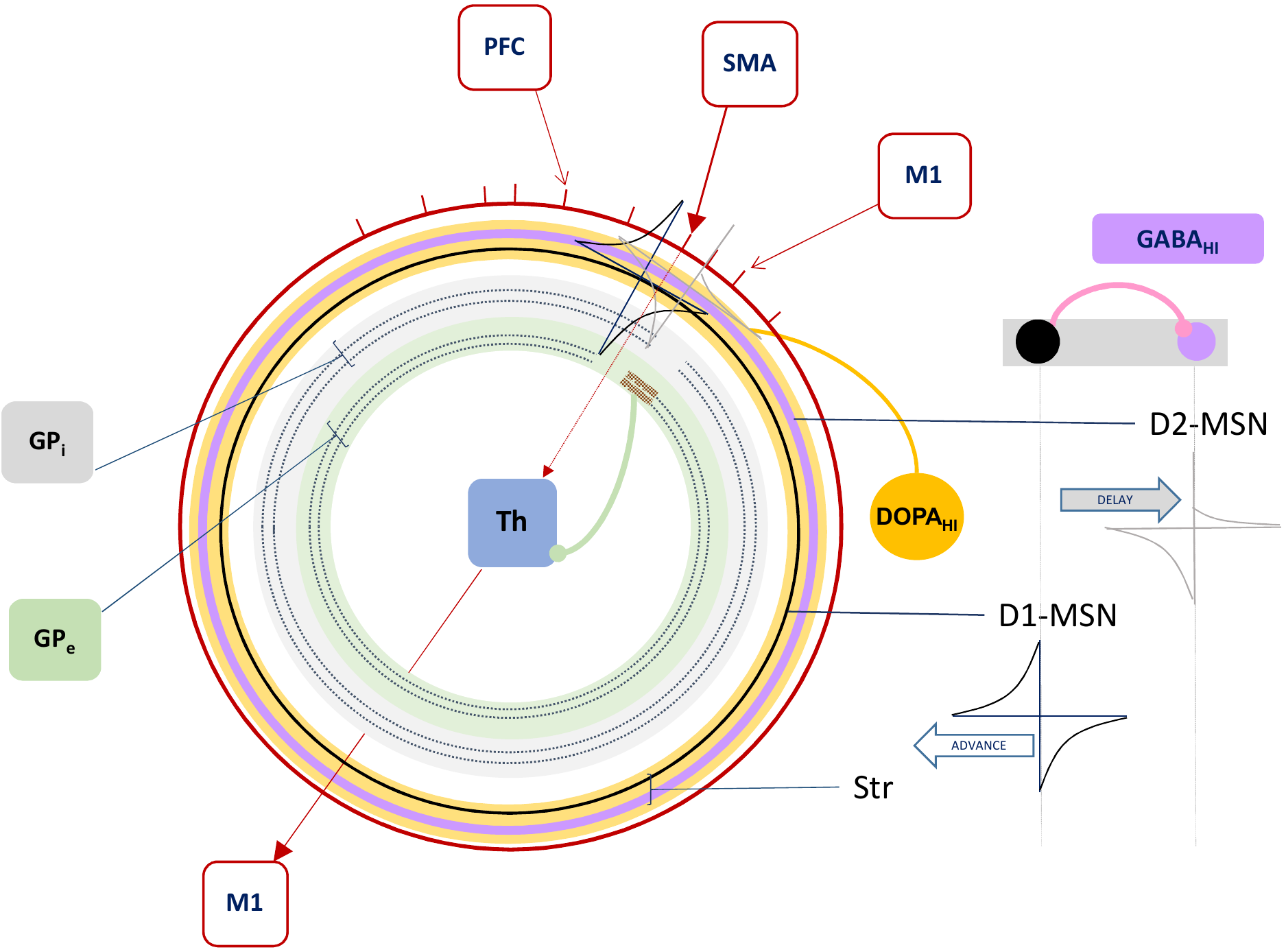}
\end{center}
\textbf{\refstepcounter{figure}\label{FDG_Loop} Figure \arabic{figure}.}{ The Forward Driver Gate. Cortical action potentials (red raster marks) traverse the Grand Loop (red circle, representing a periodic time line), when neurons in specific areas (red boxes) spike. These cause spikes in striatum, represented by STDP functions placed on a periodic time line for both the indirect (lavender circle) and direct (black circle) pathway medium spiny neurons (D$2$-MSN and D$1$-MSN). The D$1$-MSN is inhibiting the D$2$-MSN providing additional GABA-ergic modulation of STDP. Spikes cause direct disinhibition of the external segment of Globus Pallidus ($\mathrm{GP{_e}}$), allowing a cortical spike to be relayed through the thalamic gate (red arrows, SMA to M1), or indirect additional inhibition through the internal segment ($\mathrm{GP{_i}}$), blocking spikes.}
\end{figure}
\newpage
\begin{figure}[h!]
\begin{center}
\includegraphics[width=15cm]{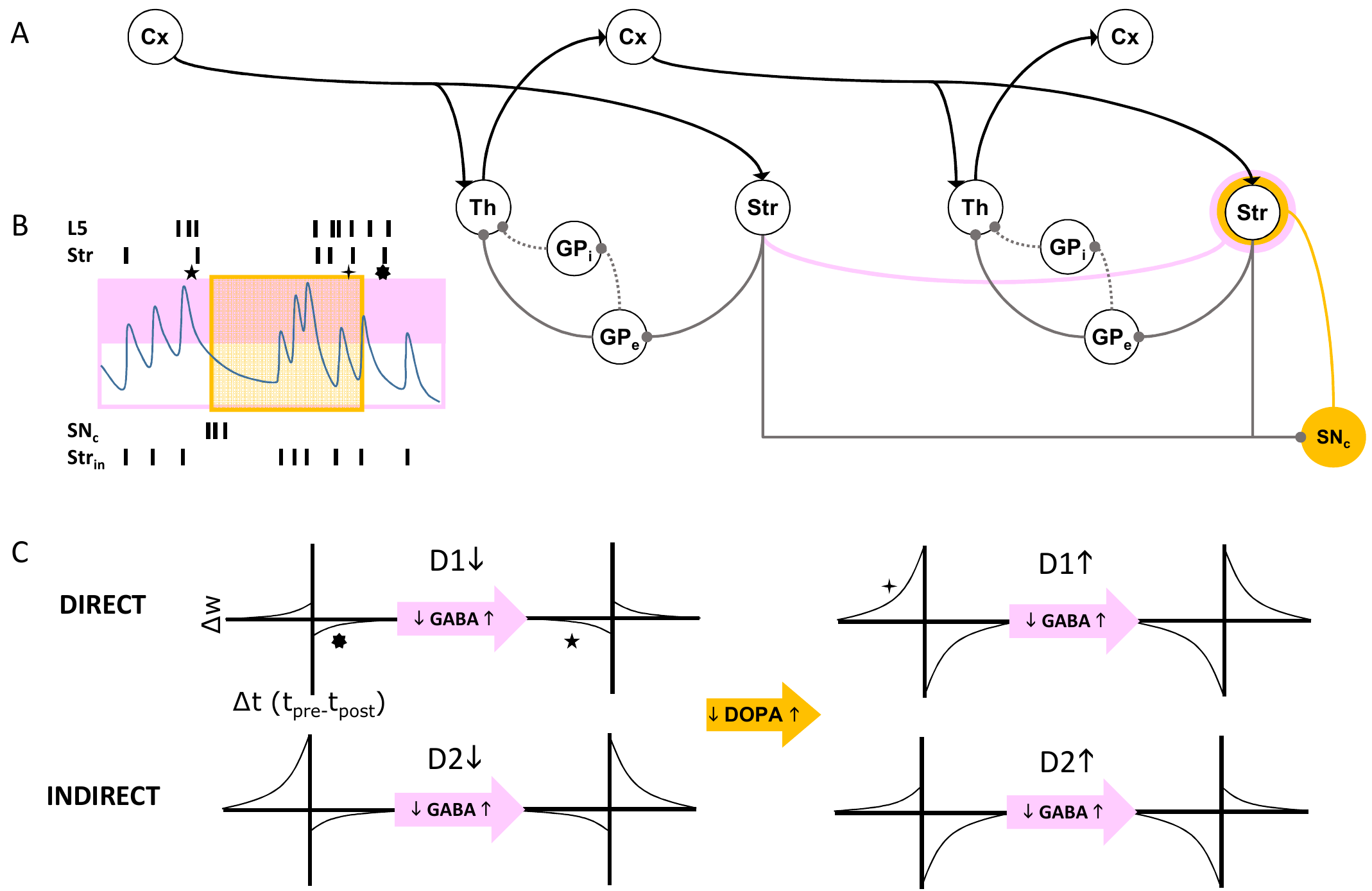}
\end{center}
\textbf{\refstepcounter{figure}\label{FDG} Figure \arabic{figure}.}{ A. Schematic of cortico-thalamo-cortical routing. Direct (solid line) and indirect (dotted line) pathways through GP disinhibit or inhibit thalamic relay neurons. The striatum is a source of self inhibition (pink line) creating GABA-ergic modulation of corticostriatal synapses (pink annulus). Dopamine neurons in SNc receive extensive inhibitory striatal inputs, and similarly modulate these synapses (yellow annulus). B. GABA levels change with intrinsic striatal firing $\mathrm{Str{_{in}}}$ and cross a threshold (pink box). GABA modulation results in changes to STDP due to correlated layer $5$ (L$5$) and striatal (Str) firing patterns. Superimposed dopamine modulation (yellow box), results in distinctly different STDP functions at a corticostriatal synapse (stars, corresponding to those in C.). C. For each combination of a direct pathway D$1$ or indirect pathway D$2$ (rows) striatal neuron's modulatory inputs, model STDP functions are represented. Transitions from low dopamine to high dopamine occurs from left half to right half of this matrix of functions. Transitions from low GABA to high GABA occurs between odd and even columns in the matrix.}
\begin{center}
\includegraphics[width=15cm]{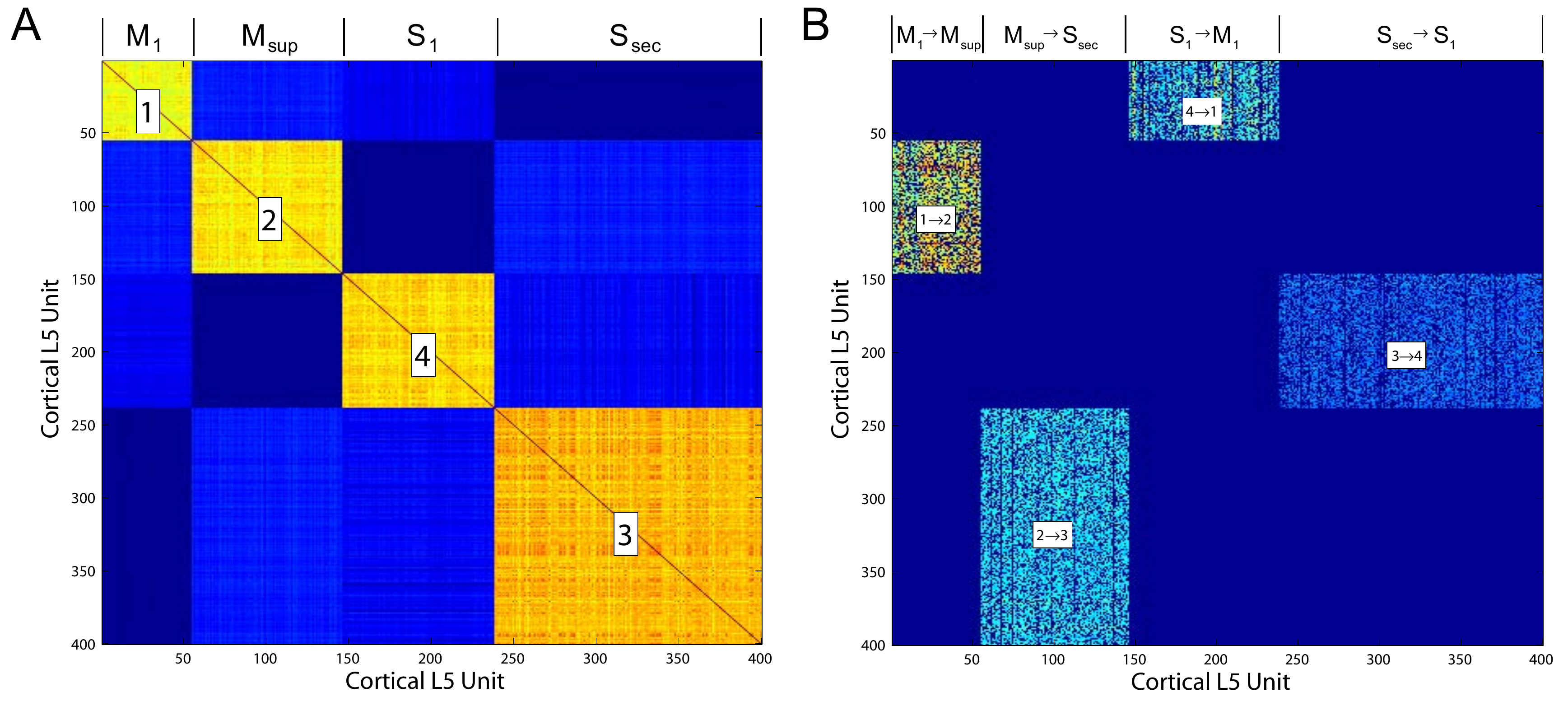}
\end{center}
\textbf{\refstepcounter{figure}\label{L5} Figure \arabic{figure}.}{ A. Correlation matrix computed over the final $10,000$ iterations of a simulation of the Layer $5$ model based on \cite{Zheng2014}. The four self-organized layers of this cortico-cortical topology are correlated in firing. B. The cortico-cortical feed back weight matrix, showing clear dominance of the feed forward area to area connections over all others. This self-organized topology supports synfire ring activity, which in the current model is referred to as traversal activity.}
\end{figure}
\newpage
\begin{figure}[h!]
\begin{center}
\includegraphics[width=12cm]{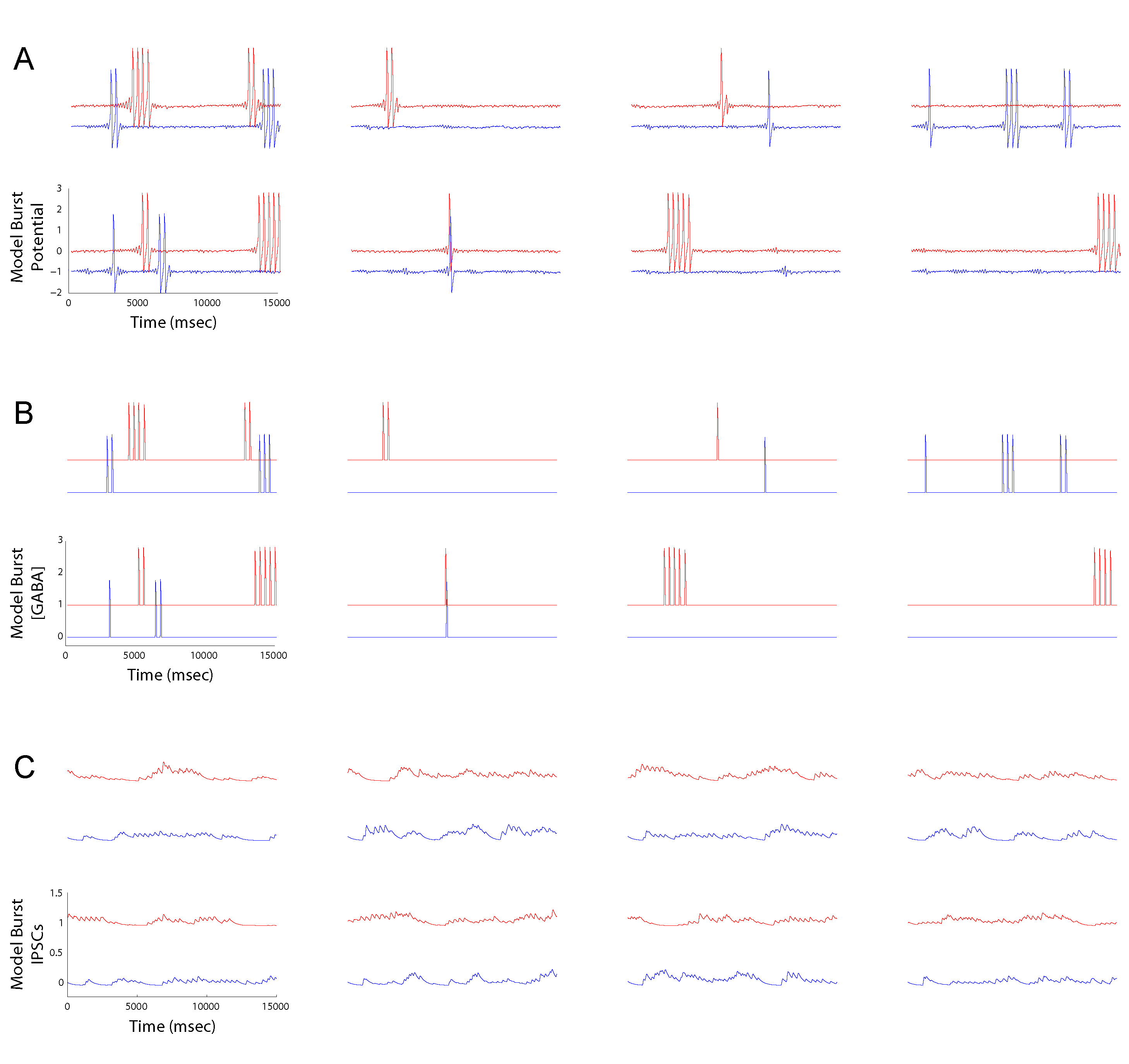}
\end{center}
\textbf{\refstepcounter{figure}\label{StrBursts} Figure \arabic{figure}.}{ A. Striatal burst potential time series derived from FitzHugh-Nagumo models of four MSNs (columns) plotted at the beginning (blue) and the end (red) of a simulation. Each burst potential represents a series of MSN spikes fired in a burst. B. The time aligned burst output traces represent the half wave rectified version of the potentials in A. C. IPSCs received by each MSN (outward currents plotted as a positive deflection). Currents are maximal when the neurons bursting ceases, and are low when it is bursting, reflecting the operation of winnerless competition in the lateral inhibitory network. }
\begin{center}
\includegraphics[width=16cm]{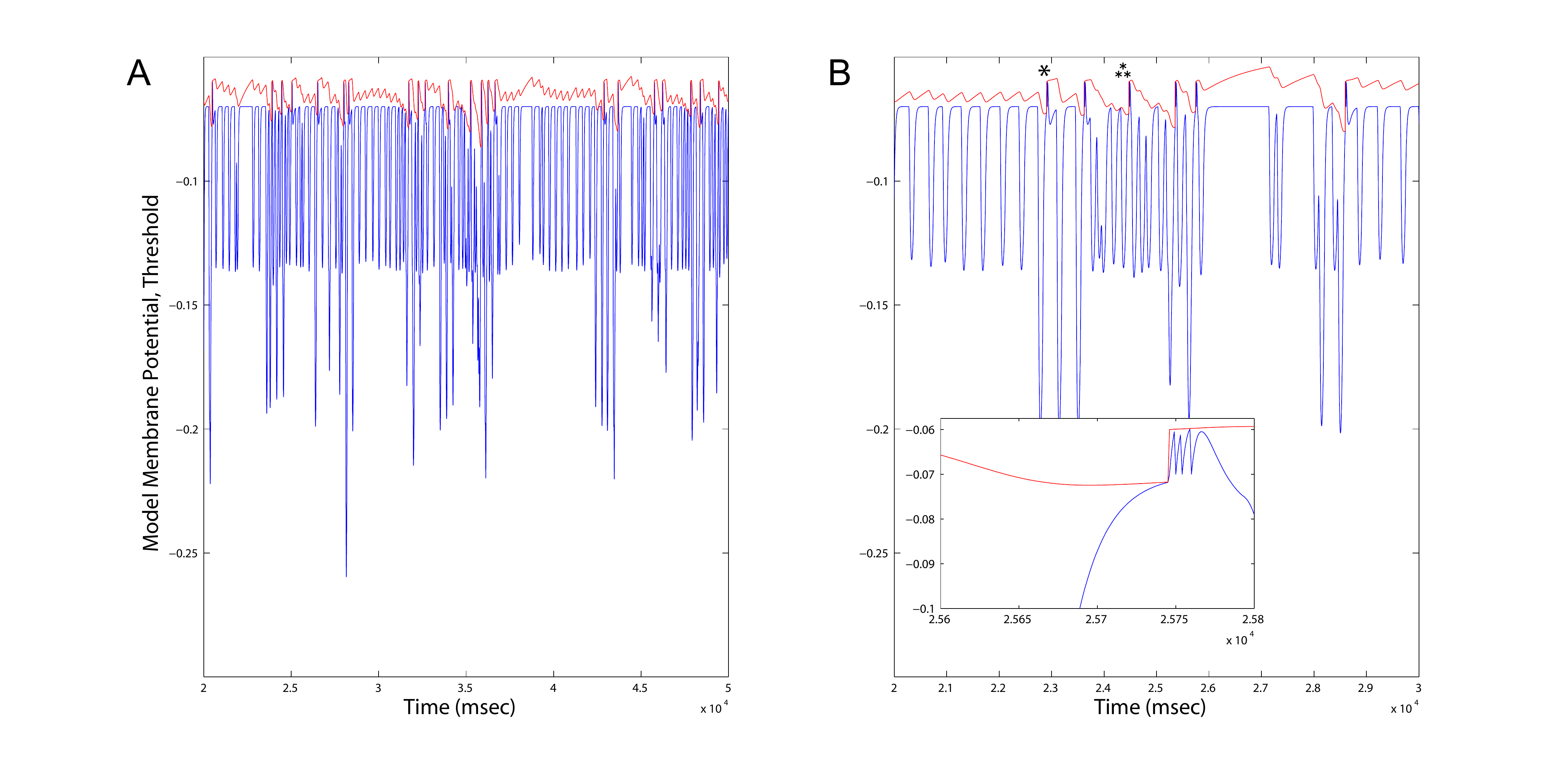}
\end{center}
\textbf{\refstepcounter{figure}\label{DA} Figure \arabic{figure}.}{ A. Long duration time series plots of Dopamine Neuron model variables. When the membrane potential (blue) reaches the variable threshold (red), a spike reset occurs. Firing rates of Dopamine neurons across all simulations were consistently on average $\sim 1.6 \mathrm{Hz}$, and varied locally depending up the ongoing integration of dynamic inputs. B. An expanded time scale reveals bursts (inset) occurring in response to deep hyperpolarization (single star), or prolonged weaker hyperpolarization (triple star) events. }
\end{figure}
\newpage
\begin{figure}[h!]
\begin{center}
\includegraphics[width=15cm]{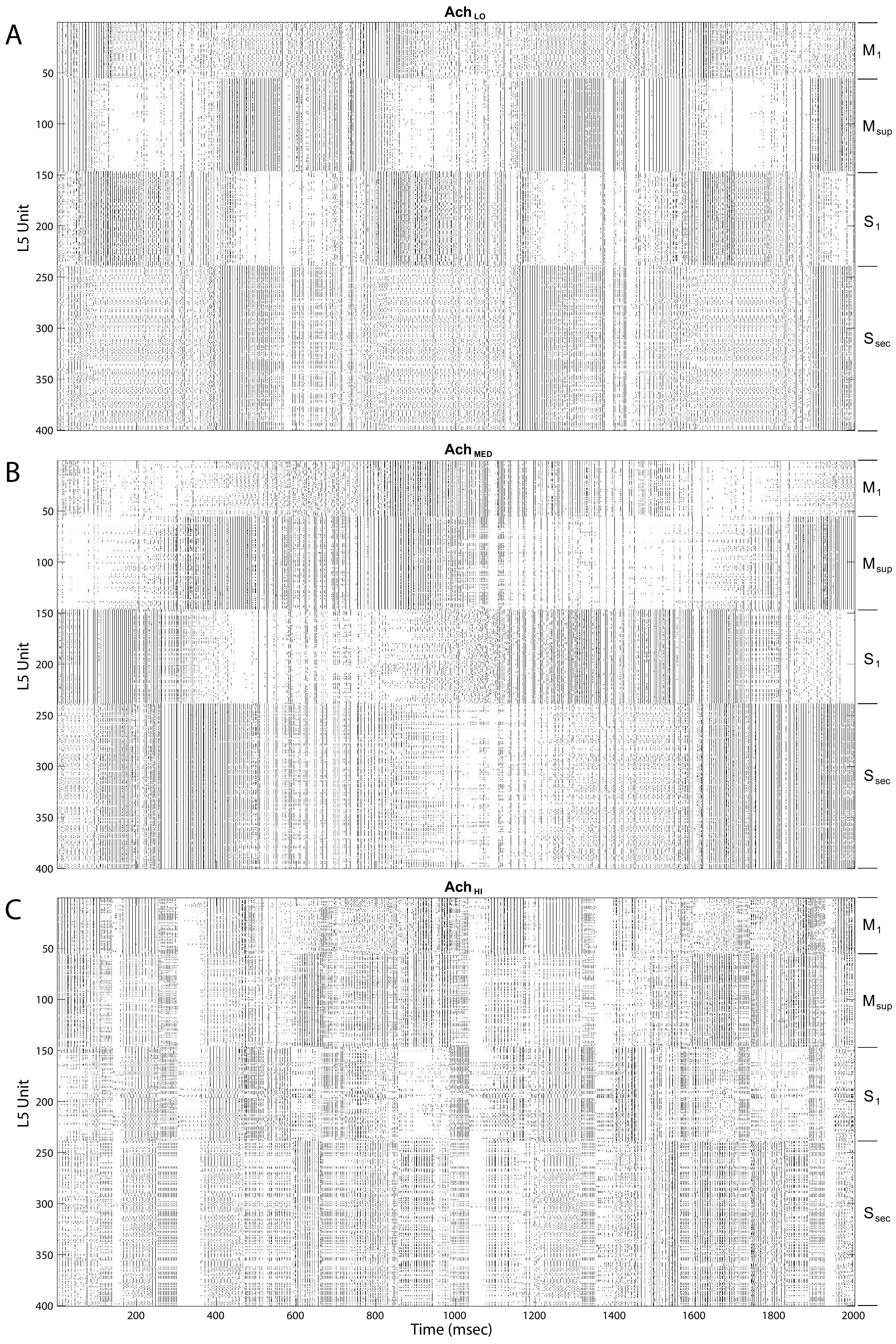}
\end{center}
\textbf{\refstepcounter{figure}\label{Raster} Figure \arabic{figure}.}{ Raster plots of the cortical Layer $5$ model outputs. $400$ spike trains from the final $2$ seconds of simulated time. Cortical areas noted on right. A. Under low $Ach$ ($0.25$) traversals are long lasting ($\sim 400$ msec) and smooth. B. Under moderate $Ach$ ($0.5$) traversals become briefer and choppy. C. Under high $Ach$ ($0.75$) traversals are brief ($100-200$ msec) and heterogenous.}
\end{figure}
\newpage
\begin{figure}[h!]
\begin{center}
\includegraphics[width=15cm]{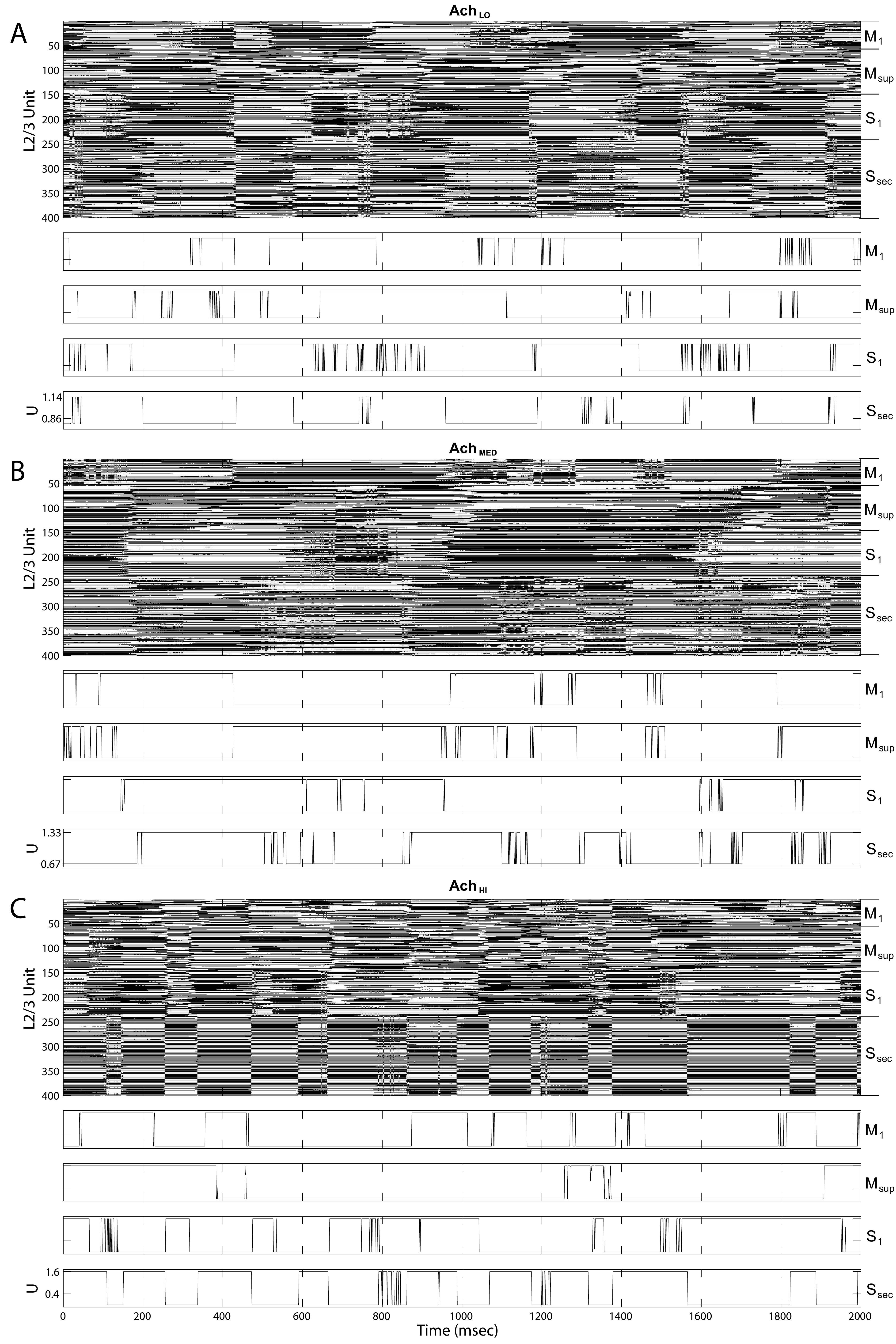}
\end{center}
\textbf{\refstepcounter{figure}\label{Up} Figure \arabic{figure}.}{ Up state (black) raster plots of Layer $2/3$ model outputs (upper panels) and example time series of gain $U$ on traversal inputs to each area (lower panels) A. Under low $Ach$ ($0.25$), states transition more quickly than traversals from Fig. \ref{Raster}. B. Under moderate $Ach$ ($0.5$), states transition more slowly in sensory areas. C. Under high $Ach$ ($0.75$), up and down states become synchronized in the secondary sensory area.}
\end{figure}
\newpage
\begin{figure}[h!]
\begin{center}
\includegraphics[width=15cm]{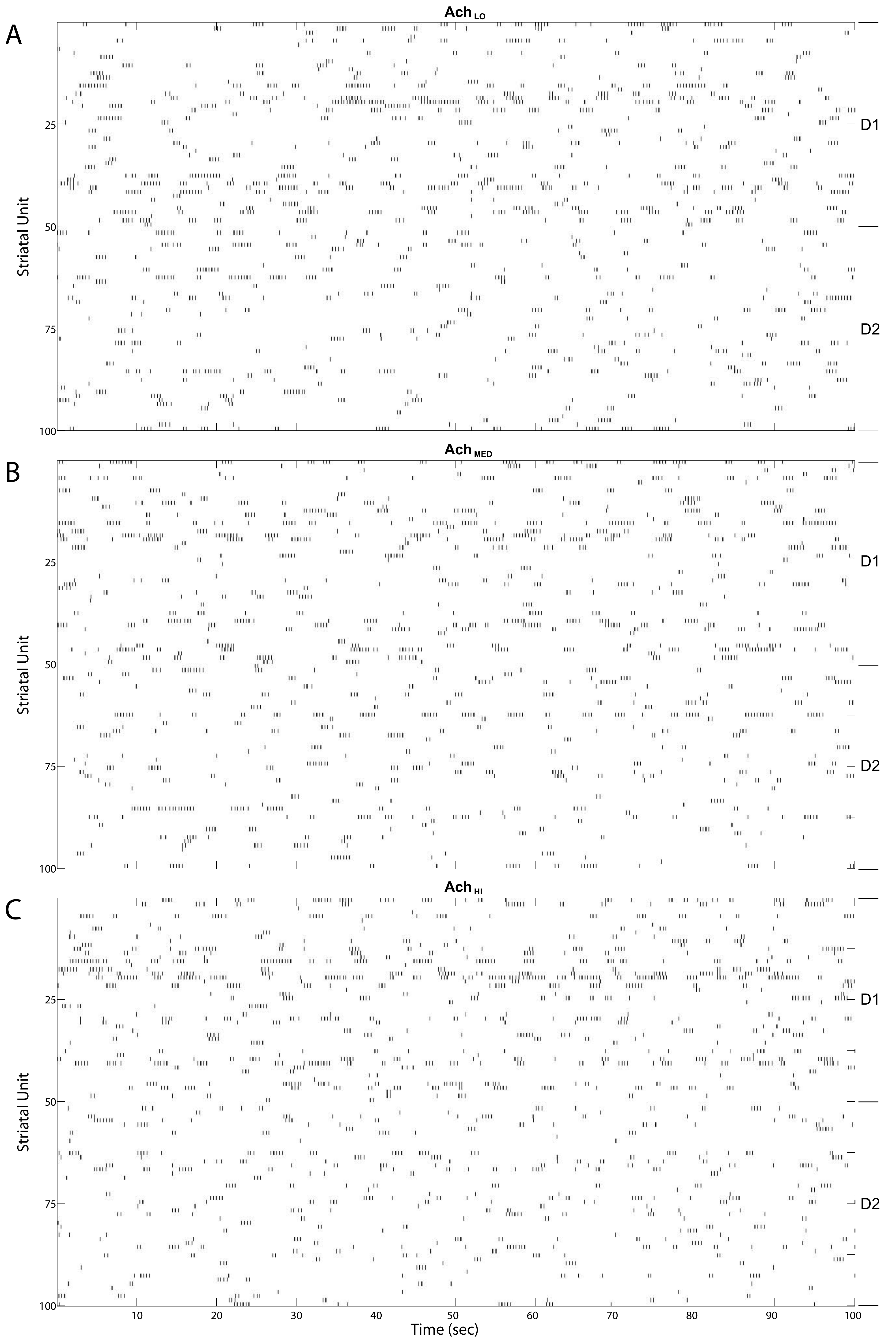}
\end{center}
\textbf{\refstepcounter{figure}\label{StrRaster} Figure \arabic{figure}.}{ Striatal model MSN burst raster plots over the final $100$ seconds of simulated time. Bursts come in alternating bursts of bursts across the population due to winnerless competition. D$1$-type and D$2$-type of MSN noted at right. Bursts of bursts are longer in duration among D$1$-type MSNs under A. Low ($0.25$), B. Moderate ($0.5$), and C. High ($0.75$) $Ach$.}
\end{figure}
\newpage
\begin{figure}[h!]
\begin{center}
\includegraphics[width=13cm]{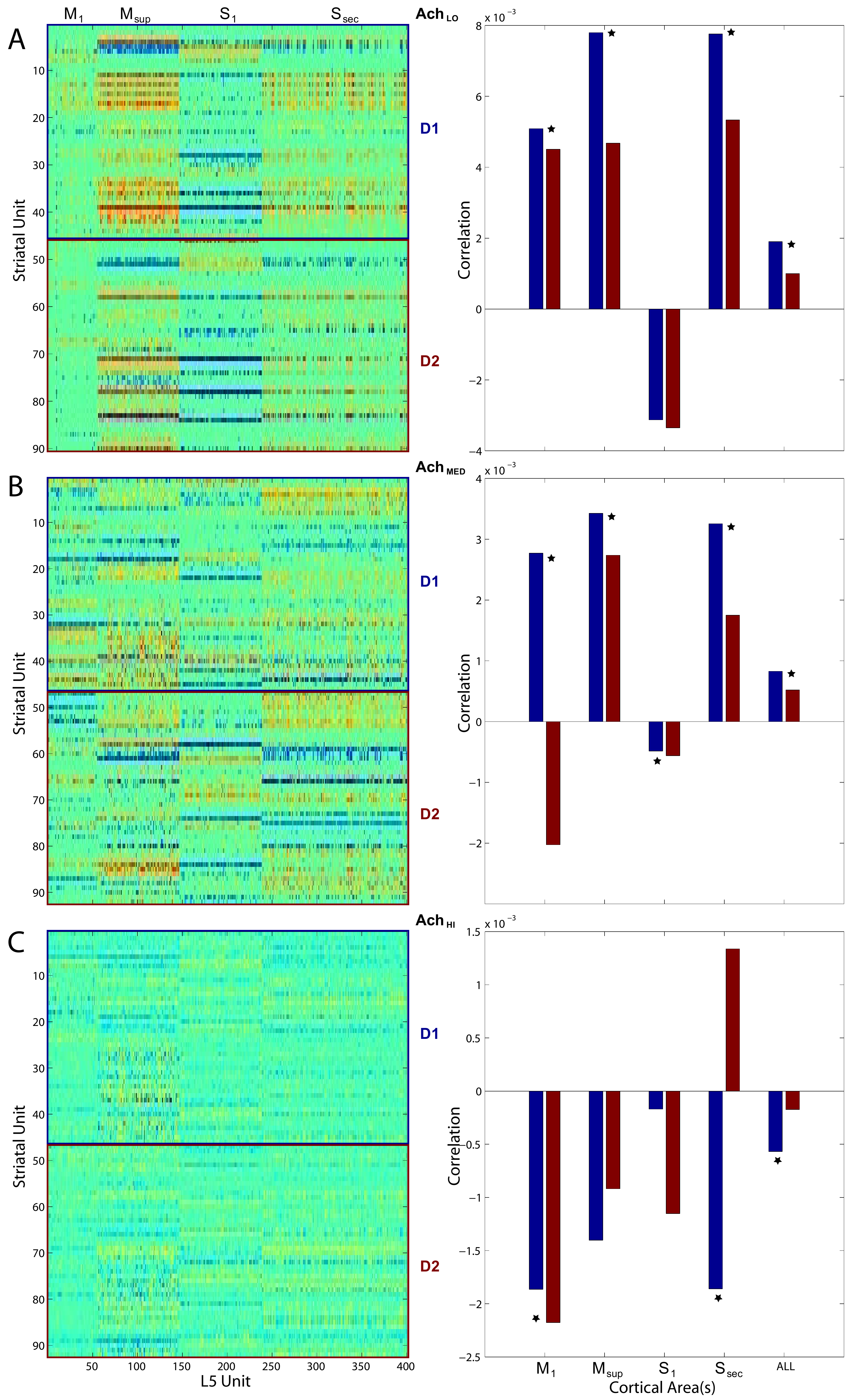}
\end{center}
\textbf{\refstepcounter{figure}\label{Corr} Figure \arabic{figure}.}{ Pairwise linear correlations coefficients between cortical spike trains and striatal bursts from Figs. \ref{Raster} and \ref{StrRaster}. Only significant correlations are shown in the matrices plotted on left (D$1$- and D$2$-types noted along right) with others shown as zero. Mean correlation values are plotted on right, for D$1$- (blue) and D$2$-type MSNs (red). A star indicates significant difference ($p<0.05$) in the distributions of coefficients (pairwise student t-test). A. Under low $Ach$ ($0.25$), correlations are positive across motor areas and all areas combined. B. Under moderate $Ach$ ($0.5$), correlations become positive for primary motor D$1$ and negative for D$2$. C. Under high $Ach$ ($0.75$), correlations are negative across motor areas and all areas combined.}
\end{figure}
\newpage
\begin{figure}[h!]
\begin{center}
\includegraphics[width=8.5cm]{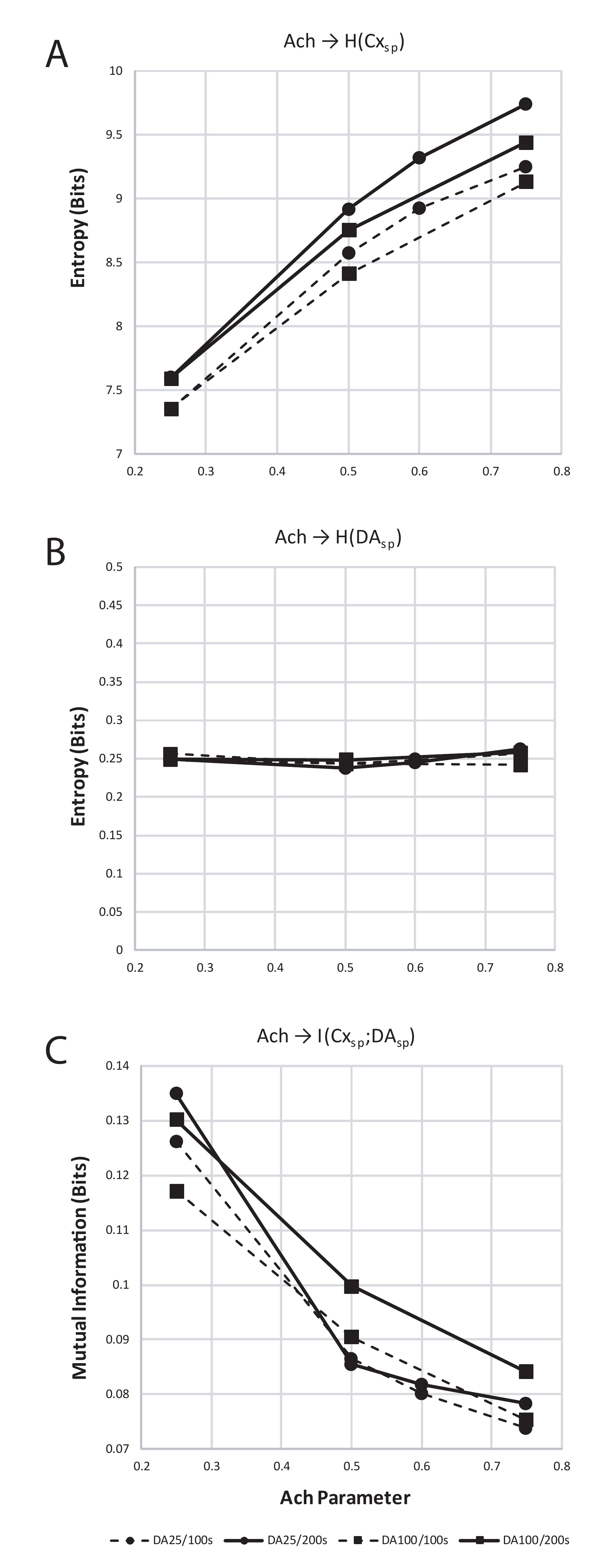}
\end{center}
\textbf{\refstepcounter{figure}\label{Ach} Figure \arabic{figure}.}{ Entropy and mutual information computed for cortical and dopamine neuron population spiking. A. Entropy ($H$) of cortical population spiking for final $100$ (dashed) and $200$ (solid line) seconds of simulated time under increasing $Ach$. Circles plot simulations with $\tau_{\mathrm{DA}}$ equal to $25$ msec and squares $100$ msec. In all conditions $H$ increases with $Ach$ and $\tau_{\mathrm{DA}}$. B. Same as A, but showing entropy of dopamine neuron population spiking. C. With increasing cortical population spiking entropy (A.), the mutual information between cortical and dopamine neuron population spiking decreased.}
\end{figure}
\newpage
\begin{figure}[h!]
\begin{center}
\includegraphics[width=15cm]{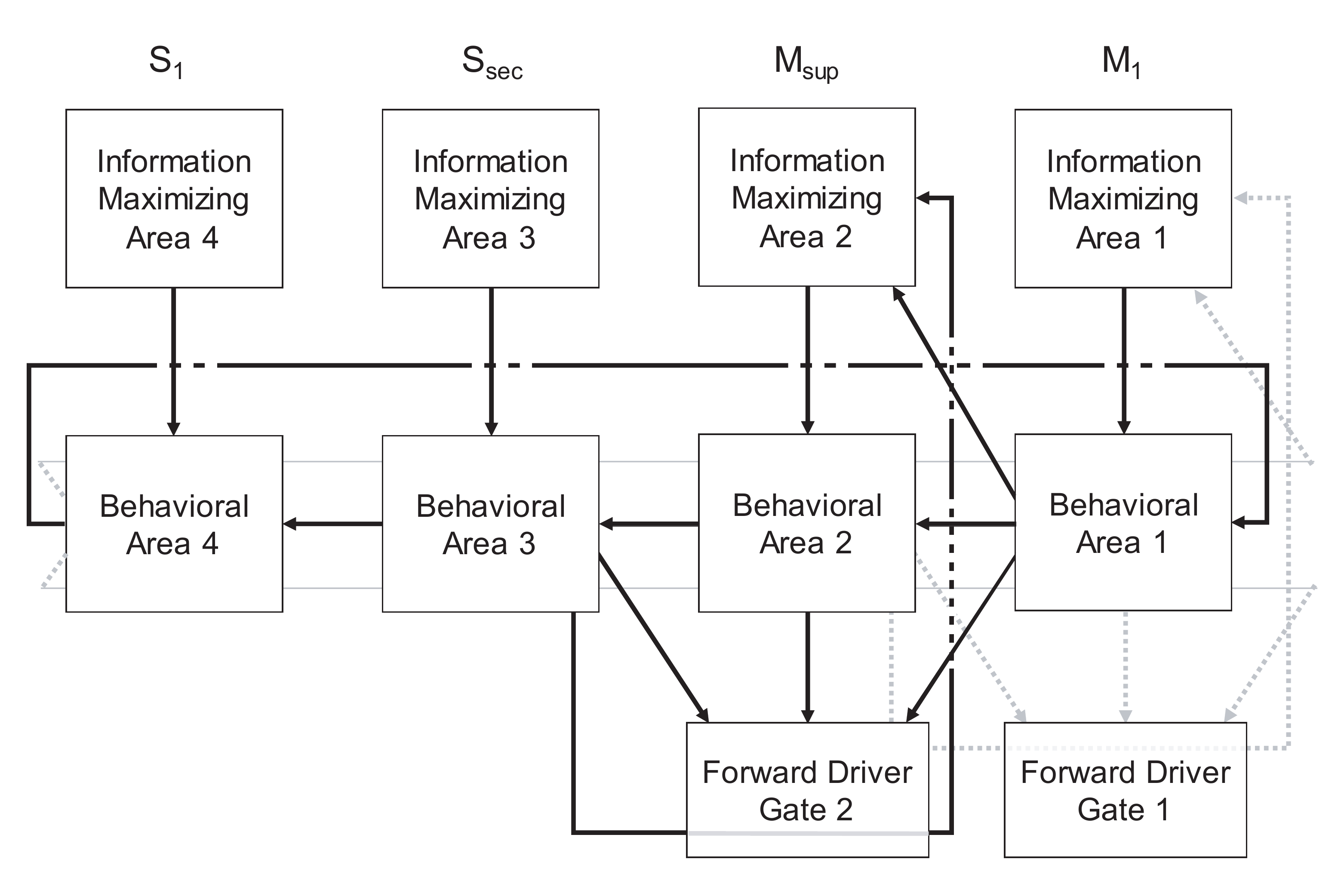}
\end{center}
\textbf{\refstepcounter{figure}\label{Map} Figure \arabic{figure}.}{ A generalized schematic for an information based exchange network. }
\end{figure}




\end{nolinenumbers}
\end{document}